\documentclass[aps,twocolumn,prd,superscriptaddress,showpacs,preprintnumbers,nofootinbib]{revtex4-1}

\usepackage{mathrsfs}

\usepackage{dsfont}
\usepackage{graphicx}
\usepackage{slashed}
\usepackage{amsmath,amssymb,amsthm,amsfonts}
\usepackage{dsfont,bbm}
\usepackage{color,rotating}
\usepackage{hyperref}
\usepackage[caption=false]{subfig}
\def\0#1#2{\frac{#1}{#2}}
\def\eq#1{(\ref{#1})}

\newcommand{\imag}{\text{i}}
\newcommand{\skipthis}[1]{}

\newcommand{\Tr}{{\text{Tr}}}

\def\s0#1#2{\mbox{\small{$ \frac{#1}{#2} $}}}
\def\0#1#2{\frac{#1}{#2}}

\newcommand{\beq}{\begin{equation}}
\newcommand{\eeq}{\end{equation}}
\newcommand{\beqa}{\begin{eqnarray}}
\newcommand{\eeqa}{\end{eqnarray}}

\graphicspath{{./figures/}}

\newcommand{\bea}{\begin{eqnarray}}
\newcommand{\eea}{\end{eqnarray}}

\def\eq#1{(\ref{#1})}

\newcommand {\apgt} {\ {\raise-.5ex\hbox{$\buildrel>\over\sim$}}\ }
\newcommand {\aplt} {\ {\raise-.5ex\hbox{$\buildrel<\over\sim$}}\ }

\def\s0#1#2{\mbox{\small{$ \frac{#1}{#2} $}}}
\def\0#1#2{\frac{#1}{#2}}

\newcommand{\be}{\begin{eqnarray}}
\newcommand{\ee}{\end{eqnarray}}

\newcommand{\rhob}{\bar\rho}

\newcommand{\dtZr}[1]{\partial_t R_k(#1)}
\newcommand{\Zk}{Z_k}
\newcommand{\Zpps}{\bar Z_{2\pi\sigma,k}}
\newcommand{\Zsss}{\bar Z_{3\sigma,k}}
\newcommand{\Zpppp}{\bar Z_{4\pi,k}}
\newcommand{\Zppss}{\bar Z_{2\pi2\sigma,k}}
\newcommand{\Zssss}{\bar Z_{4\sigma,k}}
\newcommand{\GammatwopionpR}[1]{(\Gamma^{(2)}_{\pi,k}\left(#1\right)+R_k\left(#1\right))}
\newcommand{\GammatwosigmapR}[1]{(\Gamma^{(2)}_{\sigma,k}\left(#1\right)+R_k\left(#1\right))}
\hypersetup{
	pdftitle={Self-consistent Spectral Functions in the $O(N)$ Model from the FRG},
	pdfauthor={N. Strodthoff},
	pdfkeywords={Spectral Functions} {O(N) model} {FRG},
	bookmarksopen=false,
	bookmarksnumbered=true
}

\begin{document}

\title{Self-consistent Spectral Functions in the $O(N)$ Model from the FRG}

\author{Nils Strodthoff}
\affiliation{Nuclear Science Division, Lawrence Berkeley National Laboratory, Berkeley, CA 94720, USA}


\begin{abstract}
We present the first self-consistent direct calculation of a spectral function in the framework of the Functional Renormalization Group. The study is carried out in the relativistic $O(N)$ model, where the full momentum dependence of the propagators in the complex plane as well as momentum-dependent vertices are considered. The analysis is supplemented by a comparative study of the Euclidean momentum dependence and of the complex momentum dependence on the level of spectral functions. This work lays the groundwork for the computation of full spectral functions in more complex systems.
\end{abstract}
\maketitle
\section{Introduction}
Realtime observables such as spectral functions, form factors or transport coefficients lie at the heart of 
the understanding of the dynamical properties of strongly correlated systems in general. Unfortunately these observables are hard to obtain in
Euclidean frameworks due to the necessity of performing an analytic continuation in the external momenta. Concentrating
on spectral functions in the following, the standard procedure is to numerically reconstruct the spectral function based on
given Euclidean propagator data using for example the Maximum Entropy Method (MEM) or performing a continuation based on appropriate
fitting functions such as Pad{\'e} approximants. The common problem of these procedures is however the lack of control and
consequently the large systematic errors arising from the analytic continuation. Therefore the present situation urges for a direct calculation of spectral functions.

Such an approach has been put forward in the framework of the Functional Renormalization Group (FRG) in \cite{Strodthoff:2011tz,Kamikado:2013sia}, cf.\ also \cite{Strauss:2012dg} for a
calculation of the gluon propagator at complex momenta using Dyson-Schwinger equations. The framework has been significantly generalized in \cite{Pawlowski:2015mia}, where also the connection to real time flows on a Schwinger-Keldysh contour has been established, see also \cite{Canet:2006xu,Mesterhazy:2013naa,Mesterhazy:2015uja} for recent real time applications in the FRG. Although related methods have been employed earlier in the context of solid-state physics \cite{Kyung1998,Rohe2000} only recently there has been a growing interest in such approaches \cite{Tripolt:2013jra,Tripolt:2014wra,Pawlowski:2015mia,Tripolt:2016cey,Yokota:2016tip,Kamikado:2016chk,Jung:2016yxl}.

In this work we present a proof-of-principle study for the calculation of spectral functions in a general setting considering
the full momentum dependence of the propagators as well as momentum-dependent vertices. The model of choice for this study 
is the relativistic $O(N)$ model at vanishing temperature. While we have in particular application in QCD in mind, where the $O(4)$ model
in 3+1 dimensions represents a low-energy effective model for two-flavor QCD, relativistic $O(N)$ models mainly in 2+1 dimensions 
also play an important role as effective descriptions of condensed-matter systems such as quantum antiferromagnets or superconductors.
Although we consider the $O(N)$ model as illustrative example, the underlying framework is of a more general nature. The
primary application we have in mind is the calculation of real time observables in the framework of the fQCD collaboration \cite{fQCD:2016-10} which aims
to provide a first-principle continuum approach to QCD, see \cite{Mitter:2014wpa,Braun:2014ata,Cyrol:2016tym}. The primary objective is the calculation of the elementary spectral functions
of QCD in this framework, which can then be used as direct input for the computation of transport coefficients along the lines of \cite{Haas:2013hpa,Christiansen:2014ypa}. This necessitates an approach that is capable of handling the technically advanced truncation schemes that are required for quantitative accuracy in Yang-Mills (YM) theory and QCD \cite{Mitter:2014wpa,Cyrol:2016tym}, which for this reason has to be an entirely numerical procedure. In particular, this requires the use of regulator functions that depend on 4-momentum in contradistinction to the commonly used 3-momentum regulators and as well as the inclusion of fully momentum-dependent
propagators and momentum-dependent vertices.

The $O(N)$ model at vanishing and nonvanishing temperature has been studied in a wide range of different functional approaches ranging from the FRG
\cite{Wetterich:1992yh, 
Tetradis:1993ts, 
Morris:1997xj,
Litim:2002cf,
Blaizot:2006vr,
Blaizot:2006rj,
Benitez:2011xx,
Zappala:2012wh,
Kamikado:2013sia,
Nagy:2012np,
Rancon:2014cfa,
Rose:2015bma,
Pelaez:2015nsa}
to two-particle irreducible (2PI) approaches
\cite{Lenaghan:1999si,
Nemoto:1999qf,
Roder:2005vt,
Ivanov:2005bv,
Seel:2011ju,
Mao:2013gva,
Marko:2013lxa,
Marko:2015gpa}
focusing on the one hand one investigating properties of QCD/solid-state applications in an
effective model setting and, on the other hand, on critical behavior at the quantum critical point in $2+1$ dimensions
or the thermal critical behavior in $3+1$ dimensions. Most of the studies focus on solving the system of
Euclidean correlation functions in various degrees of sophistication, but only a few address the calculation
of realtime observables and, in particular, spectral functions in this model. Notable exceptions are the spectral functions
in 2+1 dimensions calculated from Euclidean data by means of Pad{\'e} approximants \cite{Rancon:2014cfa,Rose:2015bma},
finite-temperature spectral functions from the lattice \cite{Engels:2009tv} via MEM, and the first FRG calculation for spectral functions
in the present framework \cite{Kamikado:2013sia}, where spectral functions were calculated on the basis of a Euclidean solution in the local potential approximation (LPA),
as well as a 2PI investigation on finite-temperature $O(N)$ model spectral functions \cite{Roder:2005vt}. The latter uses an approximation scheme that includes similar diagrams as the truncation scheme with momentum-independent vertices considered in this work. However, the numerical solution neglects the real part of the diagrams involving 3-point diagrams which lead to a deviations from classical Euclidean propagators. The spectral functions computed in this work do not only include the full complex momentum dependence of the propagators but in addition momentum-dependent vertices
and are therefore the most sophisticated directly calculated spectral functions and at the same time the first results on the way towards quantitative elementary spectral functions in YM theory and QCD.

The paper is organized as follows: In Sec.~\ref{sec:methods} we review the $O(N)$ model in a vertex expansion scheme for the effective action focusing on the Euclidean 
momentum dependence in the first part and the direct calculation of realtime propagators in the second part. Results for Euclidean propagators and vertices as well as spectral functions in different
approximation schemes are presented in Sec.~\ref{sec:results}. We summarize and conclude in Sec.~\ref{sec:summary}.

\section{Euclidean and Realtime correlation functions from the FRG}
\label{sec:methods}
\subsection{$O(N)$ model in a vertex expansion scheme}
\label{sec:onmodel}
The classical action of the $O(N)$ model is given by
\begin{equation}
\label{eq:Scl}
S_{\text{cl}}=\int d^d x \left(\frac{1}{2}\partial_\mu \phi_a \partial_\mu \phi_a +m^2 \rho +\frac{\lambda}{2} \rho^2- c \sigma \right)\,,
\end{equation}
where $\phi_a=(\sigma, \vec\pi)_a$, $\rho(x)=\frac{1}{2}\phi_a \phi_a$ and the explicit symmetry breaking terms $c \sigma$ accounts
for a nonvanishing pion mass. Here we work in $d=4$ dimensions. 

In this work quantum fluctuations are included by means of the functional renormalization group (FRG). The FRG implements
the idea of the Wilson renormalization group, where quantum fluctuations are gradually integrated,
see \cite{Berges:2000ew,Pawlowski:2005xe,Gies:2006wv,Schaefer:2006sr,Braun:2011pp} for QCD-related reviews. 
On a technical level this is achieved by means of an infrared regulator that is added to the action which is of the form
\begin{align}
\label{eq:reg}
\Delta S_k&=\frac{1}{2}\int_p \phi^a(p) R_k(p^2) \phi^a(-p)\nonumber\\[2ex]
&=\frac{Z_k}{2}\int_p \phi^a(p)\left(p^2+\Delta m_r^2\right)\, r\left(\frac{p^2+\Delta m_r^2}{k^2}\right)\phi^a(-p)\,,
\end{align}
with $\int_p=\int \frac{d^d p}{(2\pi)^d}$. Here we use an exponential regulator shape function, $r(x)=x^{m-1}/(e^{x^m}-1)$ and $m=2$.
The wavefunction renormalization factor $Z_k$ is taken from the pion propagator at vanishing momentum.
The need for
the shift in the argument of the regulator shape functions, $\Delta m_r^2$ and its implications 
is discussed in the following section. The scale dependence of the effective average action $\Gamma_k$,
the scale-dependent analogue of the effective action, is governed by a 1-loop flow equation \cite{Wetterich:1992yh},
\begin{equation}
\label{eq:master}
k\partial_k \Gamma_k[\phi]=\Tr \frac{k\partial_k R_k}{\Gamma^{(2)}_k+R_k}\,,
\end{equation}
where $\Gamma^{(2)}_k$ denotes a full field- and momentum-dependent (inverse) propagator. Flow
equations for $n-$point functions are derived straightforwardly from the master equation \eq{eq:master}
via functional differentiation with respect to the fields, see e.g.\ Fig.~\ref{fig:pisigmaflows} for a graphical representation of the
flow equations for the (inverse) propagators. Given suitable initial conditions $\Gamma_{\Lambda_{\text{UV}}}$
at some large initial UV scale $\Lambda_{\text{UV}}$ the set of flow equations derived from the master equation \eq{eq:master} 
are integrated down to $k=0$ where the infrared regulator is eventually removed and where $\Gamma_k$ approaches the effective action $\Gamma$. 

The structure of \eq{eq:master} implies that the flow equation for a $n-$point vertex function depends in turn on up 
to $(n+2)$-point vertex functions. This leads to an infinite tower of
coupled equations that have to be truncated within appropriate nonperturbative expansion schemes
in order to obtain a finite system of coupled equations suitable for numerical applications. While
frequently derivative expansions are applied in scalar models \cite{Morris:1997xj,Canet:2002gs,Canet:2003qd,Pelaez:2015nsa} such as the $O(N)$ model, we address
the $O(N)$ model in a vertex expansion in this work with particular regard to future applications
in YM theory and QCD along the lines of \cite{Mitter:2014wpa,Cyrol:2016tym}. 
The vertex expansion corresponds to an expansion of the effective action
in terms of 1PI vertex functions, see App.~\ref{app:vertexexp} for the parametrizations
of propagators and vertices used in this work.

\begin{table}[!ht]
  \centering
  \begin{tabular}{l|l}
    truncation & considered dressing functions \\
    \hline
    LPA & $U_k$ \\
    LPA'& $U_k$, $Z_{\pi,k}$ \\
    LPA'+Y& $U_k$, $Z_{\pi,k}$, $Z_{\sigma,k}$ \\
    PMOM & $U_k$, $Z_{\pi,k}(p)$, $Z_{\sigma,k}(p)$ \\
    PVMOM & $U_k$, $Z_{\pi,k}(p)$, $Z_{\sigma,k}(p)$,\\
    & $Z_{xxy,k}(p_{\text{sym}})$, $Z_{xxyy,k}(p_{\text{sym}})$
  \end{tabular}
  \caption{Overview over the considered truncation schemes, see text for details.}
  \label{tab:trunc}
\end{table}

Here we consider five different truncations:
LPA, where only a scale-dependent effective potential is considered, LPA' and LPA'+Y which add scale-dependent wavefunction
renormalization factor(s), a truncation with fully momentum-dependent propagators (PMOM) analogous to the bosonic sector in \cite{Helmboldt:2014iya}, 
and finally a truncation scheme with fully momentum-dependent propagators and momentum-dependent vertices (PVMOM). In LPA' we consider just
a single scale-dependent wavefunction renormalization constant whereas in LPA'+Y separate wavefunction renormalization factors for each field are taken into account. 
The latter can be understood as taking into account contributions to the propagators corresponding to a term of the form
$Y_k \partial_\mu \rho \partial_\mu \rho$ on the level of the effective action \cite{Wetterich:1991be}, where one neglects $Y_k$-dependent, i.e.\ momentum-dependent, 
contributions to the vertices as argued in \cite{Helmboldt:2014iya}.
The different truncation schemes are summarized in Tab.~\ref{tab:trunc}.

We solve the full system at a fixed expansion point in field space using
a Taylor expansion at a fixed bare expansion point \cite{Pawlowski:2014zaa}.
Here this concerns in particular the effective potential which is considered to capture effects beyond
the vertices up to fourth order in the vertex expansion that are calculated explicitly. For
notational simplicity we suppress the dependence on the expansion point in the following. In the PVMOM scheme we
explicitly take into account the full momentum dependence of the propagators, and the momentum dependence of
all 3- and 4-point vertices in a one-dimensional momentum approximation at the symmetric point. Higher $n$-point functions 
($n>4$) are approximated momentum independently via the effective potential. Flow equations for $n$-point correlation
functions can either be derived from the effective potential or from the corresponding flow equations for the
vertices. Since all equations are derived from a single generating object, the effective action, these have to
coincide in the full theory, but deviate for approximate solutions. As the effective potential equation is less
sensitive to momentum-dependent approximations, we always decompose $n-$point functions as
\begin{equation}
\label{eq:decomposition}
\Gamma_{i,k}^{(n)}(p_1,\ldots,p_{n-1})=\Gamma_{i;0,k}^{(n)}+\Delta \Gamma_{i,k}^{(n)}(p_1,\ldots,p_{n-1})\,,
\end{equation}
where the flow of $\Gamma^{(n)}_{i;0,k}=\Gamma^{(n)}_{i,k}(0,\ldots,0)$ is calculated from the effective potential equation and the
momentum-dependent difference $\Delta \Gamma^{(n)}_{i,k}(p_1,\ldots,p_{n-1})$ is calculated from the corresponding
propagator/vertex equation. For $n>4$ we neglect the momentum dependence and consider just the contribution
from the effective potential. 

In summary, the numerical solution in the PVMOM truncation involves the solution of the flow equation for the effective potential $U_k$, expanded up to
some finite order $n$ in $\rho$, which also enters the momentum-independent contributions in the propagators
and vertices $\Gamma_{i;0,k}^{(n)}$ in \eq{eq:decomposition}. The numerical results below were obtained using an 
expansion order $n=7$. In the Euclidean sector it involves the simultaneous solution of 
the equations for momentum-dependent propagators $\Delta \Gamma_{\pi,k}^{(2)}$, $\Delta \Gamma_{\sigma,k}^{(2)}$ as
that of momentum-dependent 3- and 4-point vertices, $\Delta \Gamma_{2\pi\sigma,k}^{(3)}$, $\Delta \Gamma_{3\sigma,k}^{(3)}$,
$\Delta \Gamma_{4\pi,k}^{(4)}$, $\Delta \Gamma_{2\pi2\sigma,k}^{(4)}$, $\Delta \Gamma_{4\sigma,k}^{(4)}$. The momentum
dependence of all vertex function is approximated using a one-dimensional momentum approximation at the symmetric point $p^2_{\text{sym}}=\frac{1}{n}(p_1^2+\ldots p_n^2)$. The PMOM truncation neglects the momentum dependence of the vertices by setting $\Delta \Gamma^{(3)}_{x,k}=\Delta \Gamma^{(4)}_{x,k}=0$. In the LPA'+Y truncation, the momentum dependence of the inverse propagators is in addition parametrized using only
scale-dependent dressing function i.e.\ via $\Delta \Gamma_{x,k}^{(2)}(p^2)=Z_k \bar Z_{x,k} p^2$, where $\bar Z_{x,k}$ is calculated from the propagator equation at vanishing momentum. 
In LPA' in addition $\bar Z_{\sigma,k}$ is approximated by $\bar Z_{\pi,k}$. Finally the LPA involves just a scale-dependent effective potential and trivial inverse propagators, $\Delta \Gamma_{x,k}^{(2)}(p^2)\equiv p^2$.

Despite the simplicity of the model under consideration this represents already a considerably large system of equations that
is best dealt with using appropriate tools. This work relies on the fQCD-collaboration workflow that is only
briefly recapitulated at this point. Flow equations were derived using \textit{DoFun} \cite{Huber:2011qr}, traced using
\textit{FormTracer} \cite{Cyrol:2016zqb} that makes use of FORM \cite{Kuipers:2012rf}, converted into compilable 
kernels using \textit{CreateKernels} and solved numerically using the
\textit{frgsolver}, a flexible,
object-orientated, parallel C++ framework for the solution of flow
equations. For explicit flow equations and further details on the solution procedure we refer the reader to App.~\ref{app:flows}.

The UV parameters $m^2_{\text{UV}},\lambda_{\text{UV}}$ and $c$, or equivalently the bare expansion point, are tuned to reproduce 
physical parameter values of $f_\pi=93$~MeV, associated with the minimum of the effective potential, 
the pion (curvature) mass $m_\pi=138$~MeV and the sigma meson (curvature) mass $m_\sigma$ in the IR. It is beyond the scope of this 
study to systematically investigate the effect of the sigma mass on the results. Therefore we fix a sigma mass in the
full truncation (PVMOM) and adjust the UV parameters in all other approximation schemes to match this value in addition
to physical parameter values for $f_\pi$ and $m_\pi$. The 3- and 4-point dressings are initialized momentum independently at the UV scale.

\subsection{Spectral Functions from the FRG}
\label{sec:spectral}
In this section, we briefly recapitulate the framework for the calculation of spectral functions in the general framework put forward in \cite{Pawlowski:2015mia}, its generalization to fully momentum-dependent propagators and vertices as well as the specific application in the $O(N)$ model.

The central idea for the direct calculation of spectral functions in this framework is to carry the analytic continuation on the level of the functional equation itself. This poses the complication of obtaining the correct analytic continuation, as in particular at finite temperature there is an infinite number of analytic continuations but only one with the correct asymptotic behavior. The continuation is simple to identify in the situation where the right-hand side is given as an explicit analytic expression in terms of bosonic or fermionic occupation numbers $n_b$ or $n_f$. Here one simply exploits the periodicity in the external frequency $p_0$, e.g.\ $n_b(q_0+p_0)\to n_b(q_0)$, in order to obtain the correct analytic continuation. Such an analytic form is no longer available in more complicated situations and one
has to resort to other ways of obtaining the correctly continued correlation functions from those that are most easily computable in Euclidean approaches.

\begin{figure}[t]
  \centering 
\includegraphics[width=0.4\textwidth]{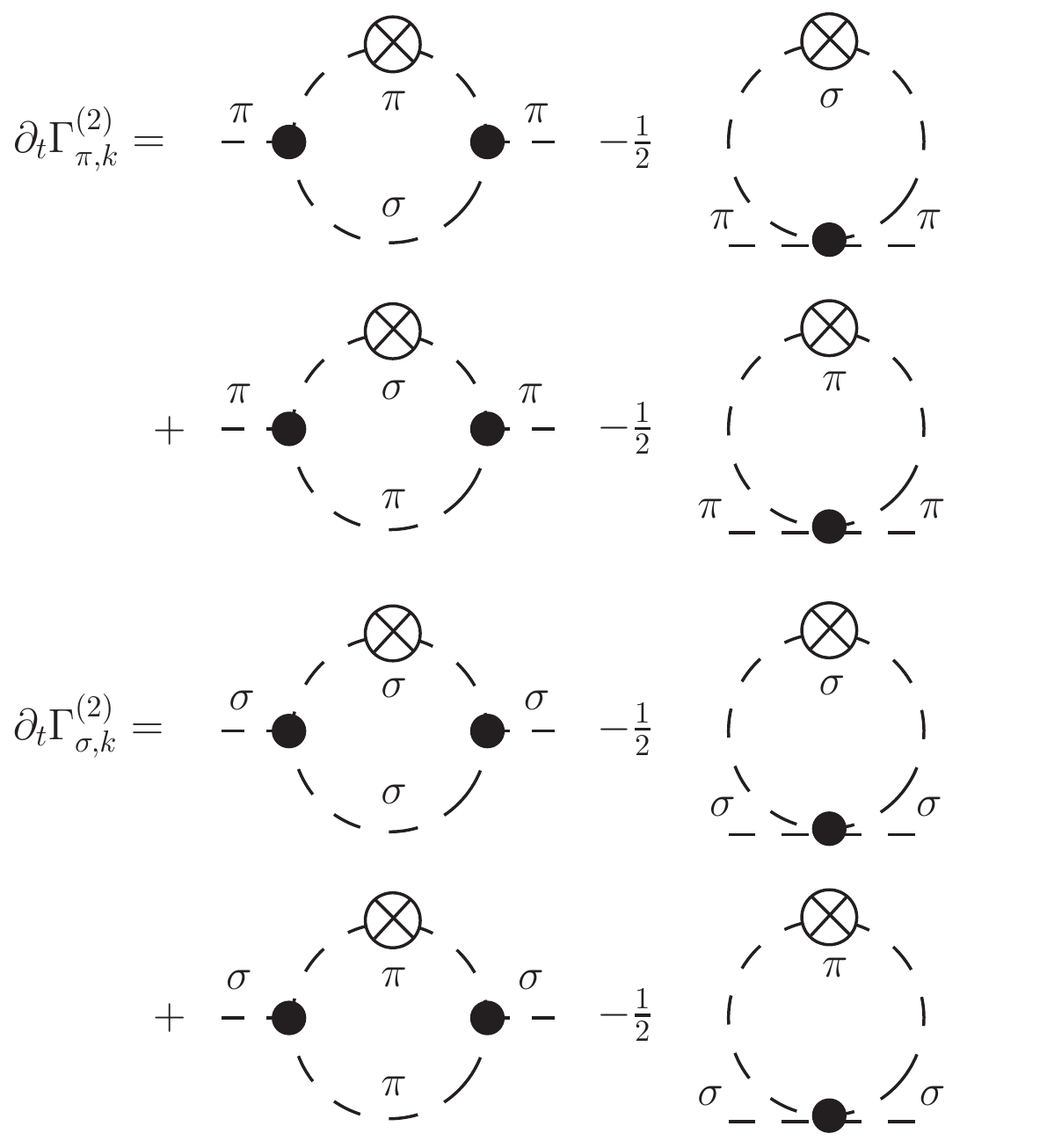}
\caption{Graphical representations of the flow equations for the inverse pion and sigma propagators. Dashed lines denote full mesonic propagators, filled circles full vertices and crossed circles correspond to insertions of $\partial_t R_k=k\partial_k R_k$. \hfill\textcolor{white}{.} }\label{fig:pisigmaflows}
\end{figure}

In the following we restrict ourselves to the case of vanishing temperature. The generalization to finite temperature is possible along the lines of \cite{Pawlowski:2015mia} and will be discussed elsewhere. The key step in this argument is to identify the difference between the
simply continued equation obtained by evaluating the right-hand side for complex momentum variables and the (inverse) retarded 2-point function, which in turn relates to the spectral function via
\begin{align}
\rho_{\sigma/\pi}(p_0,\vec p)&=-2\,\text{Im}\, G_{R\, \sigma/\pi}(p_0,\vec p)\,,
\label{eq:spectraldef}
\end{align} 
using the conventions from \cite{Pawlowski:2015mia}. The retarded propagator $G_R$ relates to its Euclidean counterpart $G_E$ via
\begin{equation}
G_R(p_0,\vec p)=-\lim_{\epsilon\to 0}G_E(- \imag(p_0+\imag \epsilon),\vec p)\,,
\label{eq:GRandGE}
\end{equation}
see \cite{Pawlowski:2015mia} for the formal generalization to finite temperature. Therefore it remains to calculate the analytically
continued Euclidean correlator, which unfortunately does not coincide with the simply continued propagator obtained by just performing the analytic continuation on the right-hand side of its equation.

\begin{figure*}[t]
  \centering \subfloat[Contours in the complex $q_0$ plane for the simple and the retarded analytic continuation arising from $G(q)G(q+p)$: simple continuation $\mathcal{C}_{\text{simple}}$ (dark blue, solid), retarded correlation function $\mathcal{C}_{\text{ret}}$ (light blue, dashed), correction contour $\mathcal{C}_{\text{corr}}$ (green, dotted).\hfill\textcolor{white}{.}]{
\includegraphics[width=0.4\textwidth]{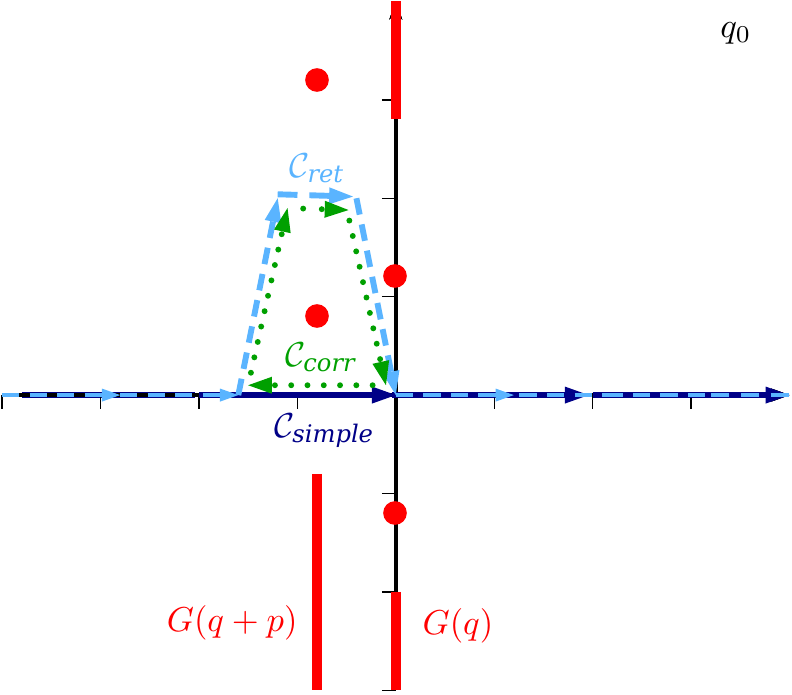}
    \label{fig:contours}} \hfill \subfloat[Constraints on $\Delta m^2_r(k)$ from avoiding regulator poles; here for $p_0^{\text{max}}=300$~MeV, $\Lambda_{\text{UV}}=900$~MeV at a mass parameter $\hat M^2=-1.3$ \cite{Pawlowski:2015mia}. A possible parametrization of $\Delta m^2_r(k)$ with parameters $\alpha=2.5$, $\beta=0.48$, and $n=20$ as in \eq{eq:Deltamr2} is shown in red. \hfill\textcolor{white}{.}]{\includegraphics[width=0.47\textwidth]{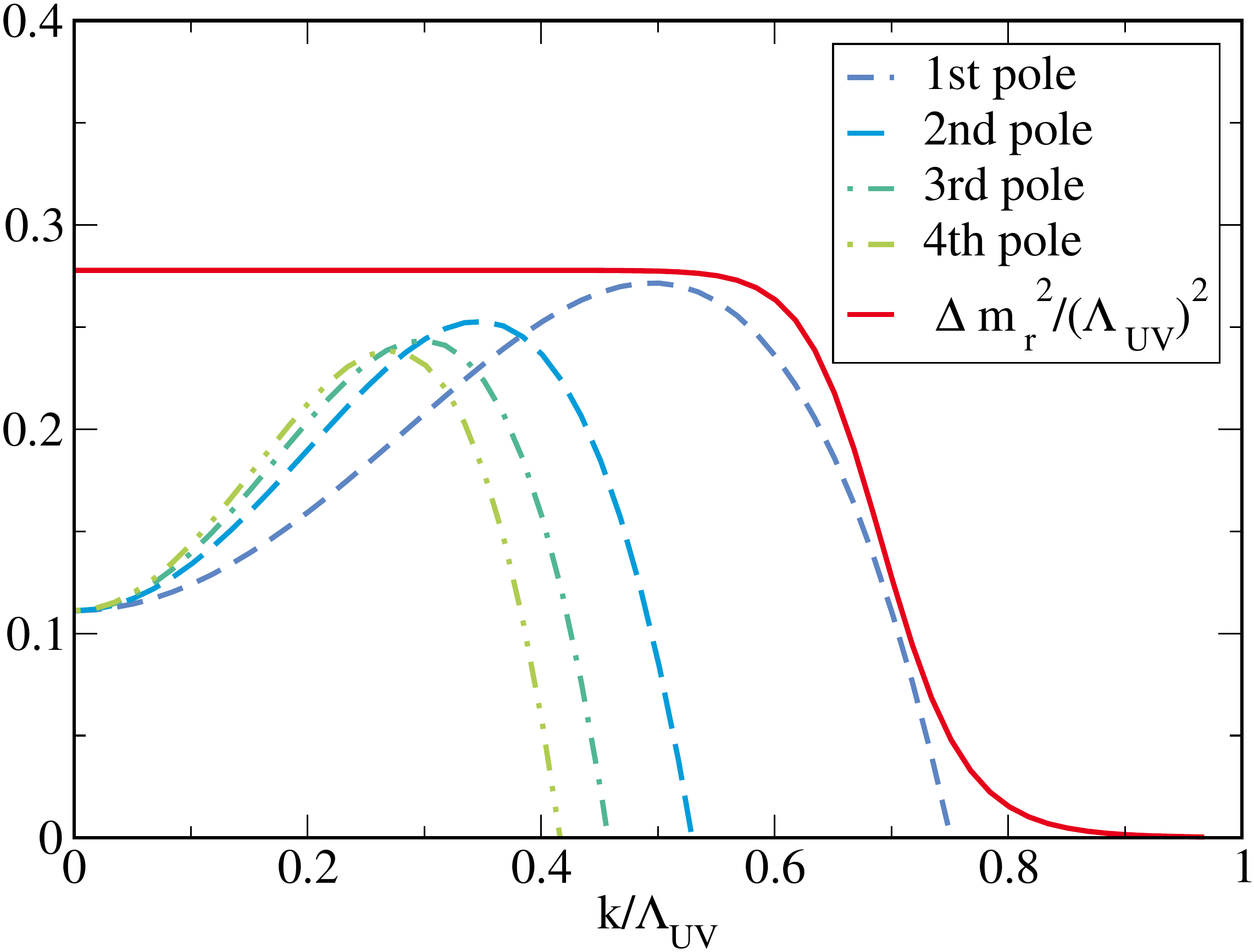}
\label{fig:poleconstraints}}
\caption{Contour choice for analytic continuation and constraints on the shift in the argument of the regulator shape function, $\Delta m^2_r(k)$ from avoiding regulator poles.\hfill\textcolor{white}{.} }\label{fig:poles}
\end{figure*}

At this point we restrict ourselves to the case of vertices that are independent of Minkowski external momentum and that depend at most on Euclidean external momentum, which implies that they can be considered as constant for the sake of this argument where we are only concerned about
the analytic structure of the right-hand side of the equation. Furthermore, one can then neglect the tadpole diagrams and consider just the diagrams
involving two 3-point vertices, see Fig.~\ref{fig:pisigmaflows}. The simplest way of understanding the difference between the two procedures is via the noninterchangeability of performing the $q_0$ integration and performing the analytic continuation. As illustrated in Fig.~\ref{fig:contours}, no matter how complicated the analytic structure of the propagator may be, the two can always be related to different integration contours. Given an analytic right-hand side of the equation in the sense of complex analysis, which is ensured by the use of analytic regulator functions cf.\ \eq{eq:reg}, the difference reduces to a closed contour integral. In the simplest case, where only simple poles are enclosed in the contour, the difference between the two continuation procedures is just given as a sum of residues, which is precisely the situation covered in \cite{Pawlowski:2015mia}. This means we can calculate the flow of the properly continued 2-point function as it appears on the right-hand side of \eq{eq:GRandGE} from the flow of the simply continued propagators and appropriate correction terms that are in the simplest case just given by residues,
\begin{equation}
\label{eq:RetardedvsEuclidean}
\Gamma_E^{(2)}(p_0^R+i p_0^I,|\vec p|)=\Gamma_{E,\text{simple}}^{(2)}(p_0^R+i p_0^I,|\vec p|)+\sum_{x\in\mathcal{C}_{\text{corr}}} \text{Res}_x\,.
\end{equation}
We stress at this point that all objects appearing on the right-hand side can be computed in purely numerical procedure which does not rely on
a certain analytic structure of the equations. Note that at vanishing temperature both continuations are identical for external momenta that are smaller than the imaginary part of the closest singularity of the propagators contributing in the diagrammatic expression for the corresponding correlation function. In case of a pole on the imaginary (i.e.\ Minkowski) momentum axis, this simply corresponds to the mass of the lightest particle. This implies, in particular, that for example the pion pole mass can already be read off from a vanishing of the simply continued (inverse) propagator.

As a final remark, calculating the spectral function from \eq{eq:spectraldef} and \eq{eq:GRandGE} requires to evaluate the continued Euclidean propagator in the limit of an infinitesimally small Euclidean external momentum. In the numerical results presented below the propagators
were however evaluated at a small but still finite value $\epsilon=0.1$~MeV for the Euclidean external momentum. In case the right-hand side of the equation can be expressed analytically in terms of occupation numbers, as it is the case for three-dimensional regulator functions, this limit can be taken analytically as demonstrated in \cite{Pawlowski:2015mia}.
This is no longer possible in the present fully numerical setup. At this point we refrain from numerically extrapolating to $\epsilon=0$ and only present results for fixed $\epsilon$ instead.

The arguments presented so far focused on 1-loop arguments without considering the occurrence of the regulator in the flow equation. As pointed out \cite{Pawlowski:2015mia}, generic analytic regulators with good numerical decay properties such as the $m=2$ exponential regulator specified below \eq{eq:reg}, lead to additional regulator poles in the regularized propagator. In this particular case there is an infinite number of regulator poles that move diagonally in the complex $q_0$ plane and approach the real axis for $k\to 0$. It has been demonstrated semi-analytically in the LPA \cite{Pawlowski:2015mia} that one can ensure that a strip around the real axis with $|\text{Im}\, q_0^\text{max}|<p_{0,\text{max}}$ stays free of regulator poles by introducing a shift $\Delta m_r^2$ in the argument of the regulator shape function in \eq{eq:reg}. Beyond LPA the absence of regulator poles in a predefined strip can no longer be proven analytically, but the position of the first pole can still be traced numerically illustrating the applicability of this mechanism in more general situations. It is advantageous to consider a scale-dependent $\Delta m_r^2=\Delta m_r^2(k)$, which is just large enough to prevent regulator poles from entering the strip. The typical $k$-dependent constraints arising from different poles are illustrated for the LPA in Fig.~\ref{fig:poleconstraints}, cf.\ \cite{Pawlowski:2015mia}. In particular, if $\Delta m_r^2(\Lambda_{UV})$ is sufficiently small such that the effective action in the UV remains unchanged compared to that with $\Delta m_r^2(\Lambda_{UV})=0$, one can continue to use the same initial conditions as for $\Delta m_r^2(\Lambda_{UV})=0$ and the results in the IR remain unchanged. In this case, the introduction of $\Delta m_r^2$ reparametrizes the flow leading to vastly different effective cutoff scales $k_\text{eff}(k)$, see \cite{Pawlowski:2015mia} for details and \cite{PSSW} for consequences in theories with different particle species. For definiteness we parametrize $\Delta m_r^2(k)$ as
\begin{equation}
\label{eq:Deltamr2}
\Delta m_r^2(k)=\frac{\alpha p_{0,\text{max}}^2}{1+\left(\frac{\beta k}{p_0^\text{max}}\right)^n}\,,
\end{equation}
for appropriately chosen parameters $\alpha=2.5$, $\beta=0.48$, and $n=20$ for $p_{0,\text{max}}=300$~MeV.

To summarize, the use of appropriate regulator functions such as the one in \eq{eq:reg} avoids the occurrence of regulator pole in a strip in the complex $q_0$ plane up to some predefined imaginary part. Provided the correction contour $\mathcal{C}_{\text{corr}}$ is entirely contained in this strip then leaves us with the 1-loop case without
regulator insertions as discussed above. In the setting of the flow equation the prescription \eq{eq:RetardedvsEuclidean} then has to be applied in every RG step. This involves tracking the poles in the propagator equations
and calculating the residue in \eq{eq:RetardedvsEuclidean}.
Here the pole search was implemented on one-dimensional slices
in the complex plane and the residue was calculated numerically via contour
integration. We stress that the restriction to simple poles
is sufficient in the given range of Minkowski momenta where only pion pole 
contributions have to be taken into account, cf.\ also the discussion on the
backcoupling effects of pole contributions in Sec.~\ref{sec:sfresults}. 
The more intricate case of larger external momenta and the numerical
treatment of possibly more complicated complex structures
in the propagators is deferred to future investigations.

In addition to the Euclidean correlation functions discussed in the previous section this requires the calculation of correlation functions
$\Gamma_{\sigma/\pi}^{(2)}(p_0^R+i p_0^I,|\vec p|)$ as a function of complex momenta i.e.\ with a three-dimensional momentum dependence.
The simply continued correlators have the convenient property that they can be calculated for fixed values of the imaginary part of the external momentum, $\text{Im}\,p_0$, as the flow equation only involves propagators $G(q+p)$ in addition to Euclidean propagators, $G(q)$. This is no longer possible if one wants to calculate the properly continued correlators via \eq{eq:RetardedvsEuclidean} that relate to the retarded correlators via \eq{eq:GRandGE}, as the evaluation of the pole corrections requires knowledge of the full three-dimensional momentum dependence of the propagator. The application of \eq{eq:RetardedvsEuclidean} in the context of the FRG leaves two possibilities: either the pole correction can be computed just on the solution of the simply continued propagators, which is the procedure that has been applied in all LPA studies so far, or the full complex propagator including pole correction can be coupled back into all complex equations, i.e.\ at the present stage in the propagator equation themselves. By construction, these procedures lead to different results as soon as a genuine imaginary part is generated during the flow. 

In the $O(N)$ model, the UV cutoff scale cannot be taken to arbitrarily large values. If we insist on keeping an essentially vanishing $\Delta m_r^2$ at the UV cutoff scale, this limits the accessible external momenta to $p_{0,\text{max}}\approx 300$~MeV, which is just large enough to identify the two-pion threshold in the sigma spectral function. This is however not a severe limitation of this approach, but just sensible from a physical point of view as the UV cutoff scale should naturally limit the accessible Minkowski external momenta. Even for the $O(N)$ model the range of external momenta could be slightly enlarged by loosening the connection to $\Delta m_r^2=0$ by simple retuning the model parameters, which would would allow us to enlarge the range in external momenta up to $p_{0,\text{max}}\approx 450$~MeV. This restriction is less severe for example in quark-meson models and eventually in the full QCD system, which admits arbitrarily large UV cutoff scales.

However, there are further decay channels at even larger external momenta that would even with an increased range in external momentum not
be accessible  as they are linked inherently to Minkowski-external-momentum dependence in vertices. One example is the three-pion threshold in the pion spectral function that arises from a momentum-dependent tadpole diagram. Its origin is most easily traced in a 2PI approach where it arises from a sunset diagram. This decay threshold is still not visible in the most sophisticated truncation PVMOM as it still neglects the Minkowski-momentum dependence in the 4-point vertex function.
As a second hint in this direction, the analysis in Sec.~\ref{sec:spectral} suggests that it is not only the propagators but also the
properly continued vertex functions that have to be consistently coupled back in the propagator equation
as well as in their own equation in order to incorporate, for example, information on resonances. This
is most easily seen at the example of a resonant 4-point vertex which we can write effectively 
as meson-exchange contribution and a residual 4-point interaction. Inserted into the tadpole diagram in the 
propagator equation this shows that the simply continued result has to fail for large external momenta
that exceed the mass of the resonance.

\section{Results}
\label{sec:results}

\subsection{Euclidean momentum dependence}
\label{sec:euclresults}

We start by comparing the five different truncation schemes discussed in Sec.~\ref{sec:onmodel} on the basis of the simplest observable
available in the setting, the minimum of the effective potential, see Tab.~\ref{tab:fpi}. 

\begin{table}[!ht]
  \centering
  \begin{tabular}{l||l|l}
    truncation & with  $\Delta m_r^2$ & without $\Delta m_r^2$\\
    \hline
    LPA & 83.1 & 83.2 \\
    LPA'& 88.8 & 88.8 \\
    LPA'+Y & 91.9 & 91.9 \\
    PMOM & 91.1 & 91.1 \\
    PVMOM & 93.0& 93.1
  \end{tabular}
  \caption{Top-down comparison of the minimum of the effective potential in different truncation schemes (in MeV)}
  \label{tab:fpi}
\end{table}

\begin{figure}[b]
  \centering 
\includegraphics[width=0.48\textwidth]{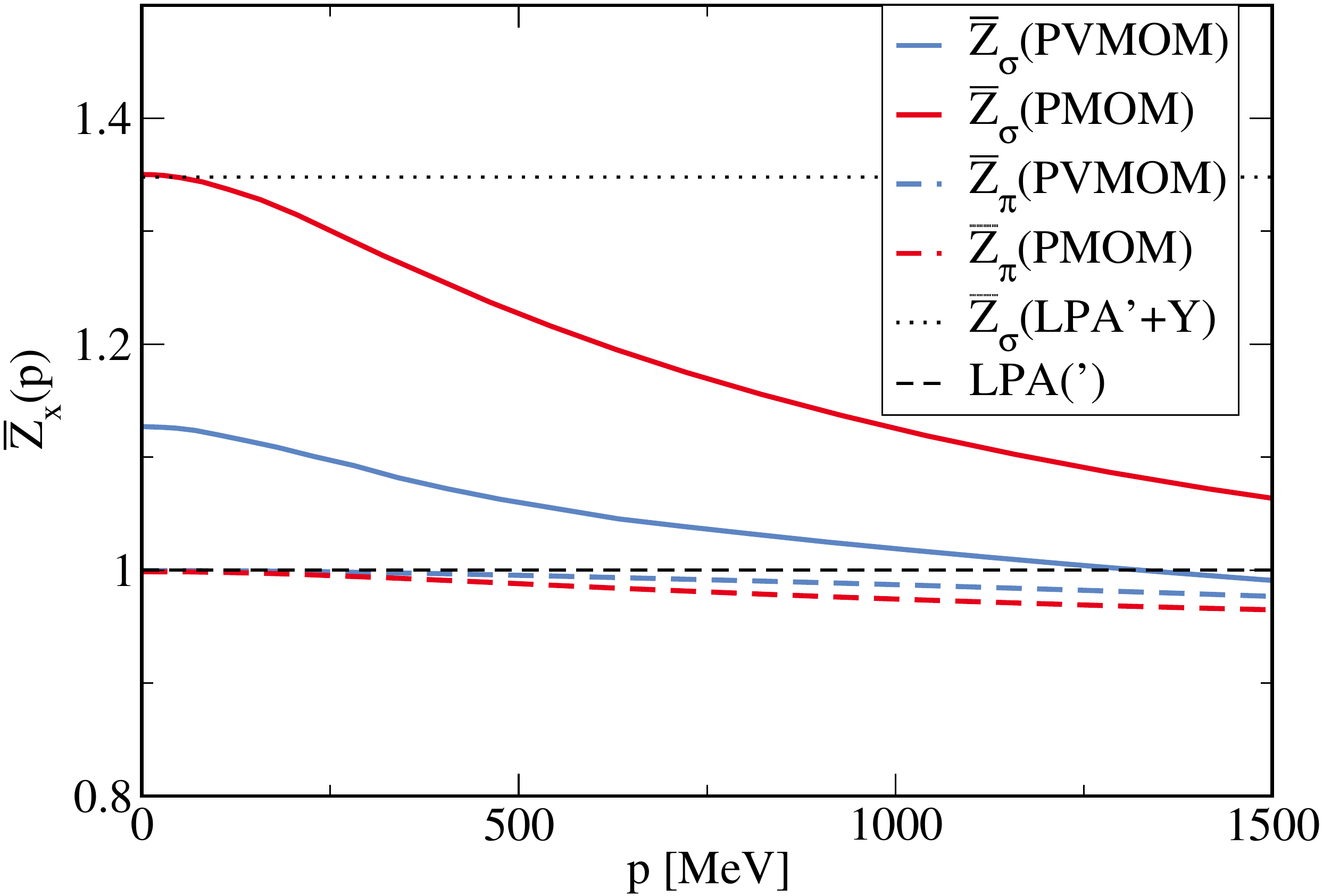}
\caption{Bottom-up comparison of the Euclidean momentum dependence of the propagators at $k=0$.\hfill\textcolor{white}{.} }\label{fig:propmom}
\end{figure}

To illustrate the differences between the different approximation schemes,
we compare the results in a top-down approach, i.e.\ in a fixed microphysics approach, where we initialize all flows at $\Lambda_{\text{UV}}=900$~MeV with the initial conditions tuned for the best truncation scheme (PVMOM). In addition we illustrate the independence of $\Delta m_r^2$ in this procedure, parametrized as in \eq{eq:Deltamr2}, by comparing to flows with $\Delta m_r^2=0$. As argued above for this parametrization of $\Delta m_r^2$ the infrared physics stays unchanged compared to $\Delta m_r^2=0$. The results improve gradually towards the PVMOM result, see Tab.~\ref{tab:fpi}. Coincidentally the LPA'+Y truncation is even closer to the full truncation PVMOM than the truncation with fully momentum-dependent propagators, PMOM. As observed in \cite{Helmboldt:2014iya}, the LPA'+Y truncation already represents a reasonable approximation of the full momentum dependence, PMOM. However, the deviations at vanishing temperature are slightly larger than the ones reported in this earlier work, where a Yukawa system was considered which is largely dominated by fermionic 1-loop contributions.

\begin{figure*}[t]
  \centering \subfloat[Momentum dependence of 3-point dressing functions\hfill\textcolor{white}{.}]{
\includegraphics[width=0.48\textwidth]{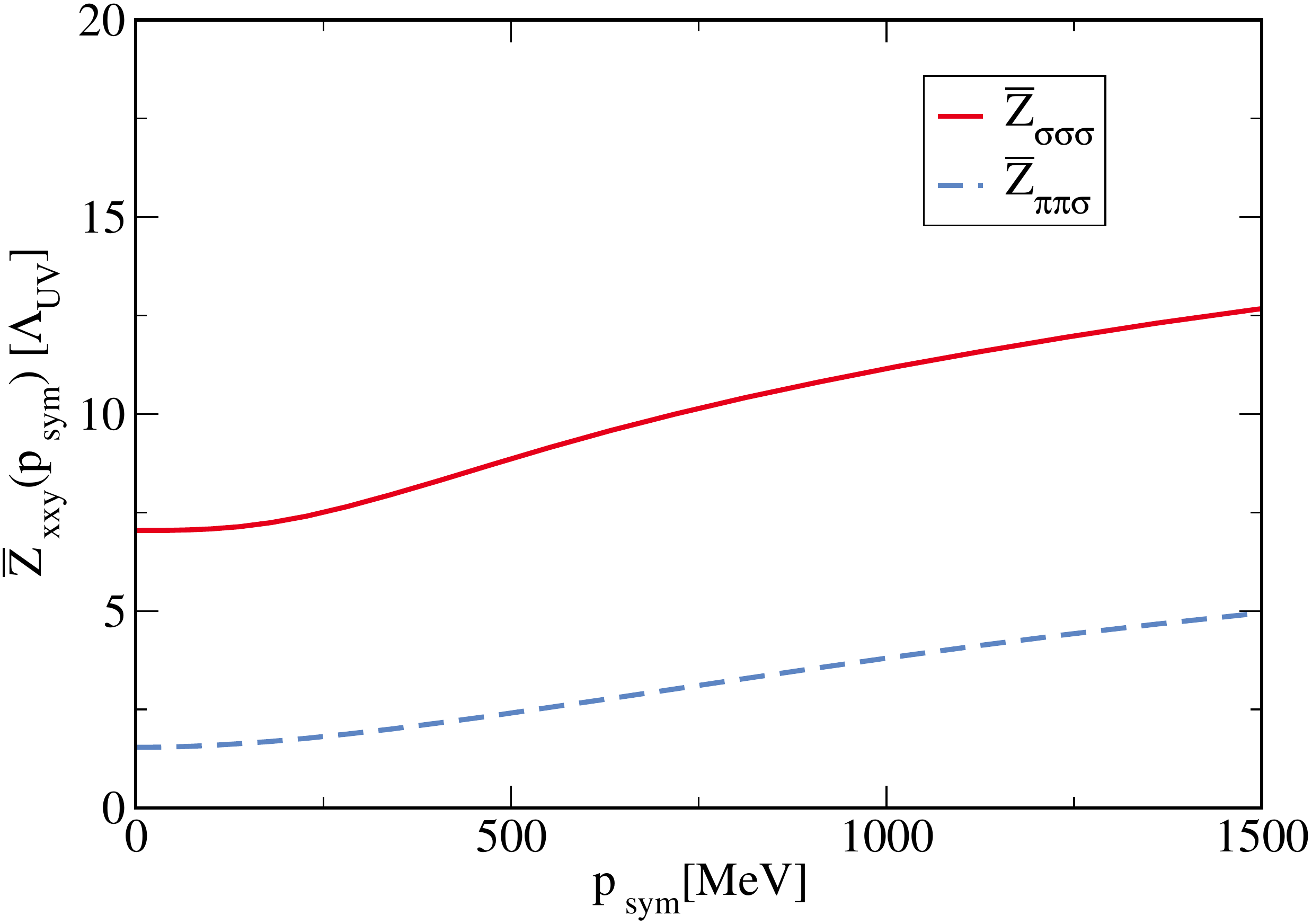}
    \label{fig:3pt}} \hfill \subfloat[Momentum dependence of 4-point dressing functions.\hfill\textcolor{white}{.}]{\includegraphics[width=0.47\textwidth]{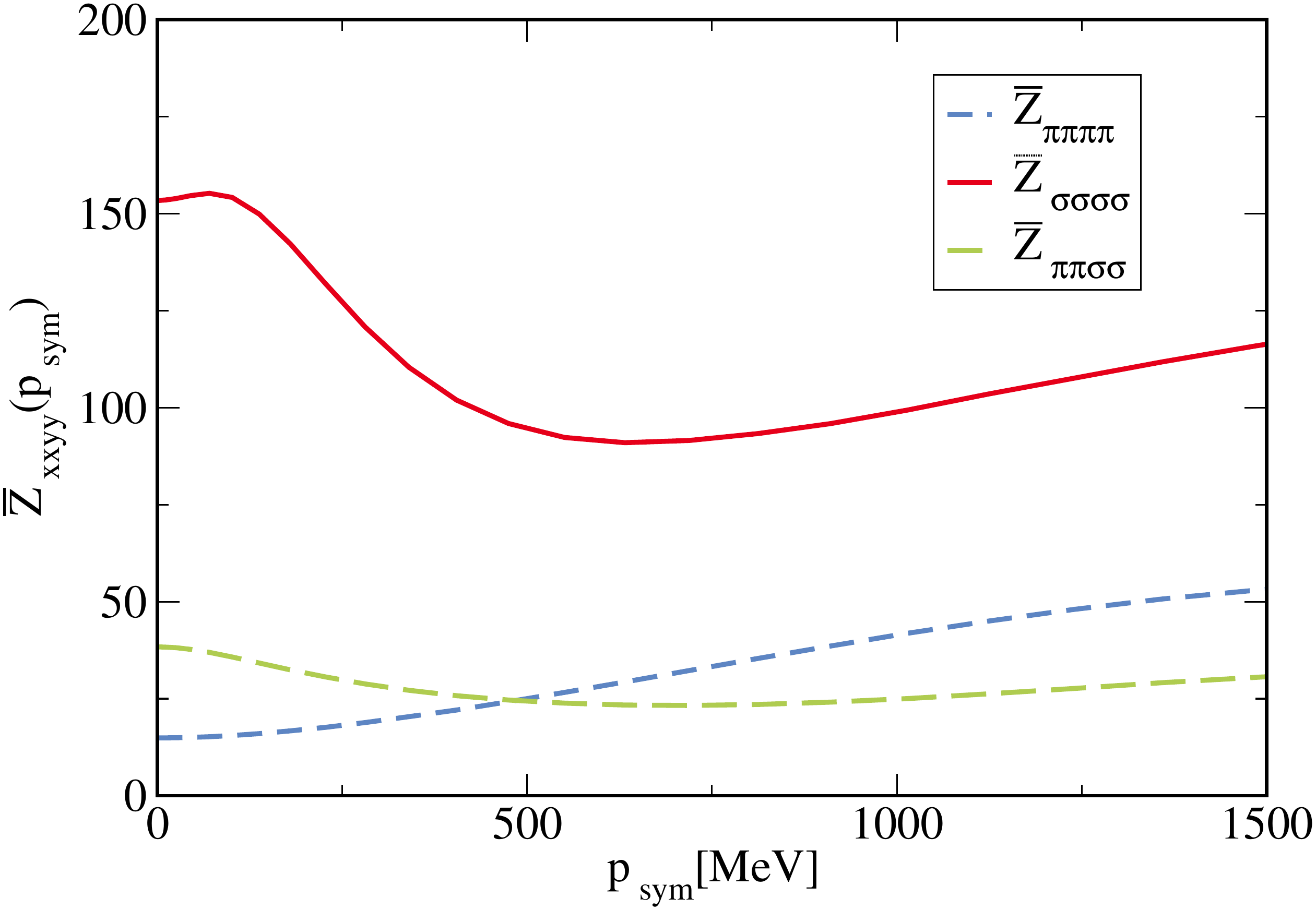}
\label{fig:4pt}}
\caption{Momentum dependence of 4- and 3- point dressing functions at the symmetric point at $k=0$.\hfill\textcolor{white}{.} }\label{fig:vertices}
\end{figure*}
Next we turn to the Euclidean momentum dependence of the propagators and vertices. In this case we choose a bottom-up approach for the comparison i.e.\ we fix the initial conditions for each truncation scheme separately as described in Sec.~\ref{sec:onmodel}. The sigma meson curvature mass was adjusted in all cases to match that of the PVMOM calculation, see Tab.~\ref{tab:mpi}. In Fig.~\ref{fig:propmom} we show the momentum dependence of the Euclidean propagators at vanishing cutoff scale, where we parametrize the (inverse) propagator via
\begin{equation}
\Gamma^{(2)}_{xx}(p)=Z (\bar Z_x(p)p^2+\bar{m}_x^2)
\end{equation}
with $\lim_{p\to 0} \bar Z_x(p)p^2=0$ and $Z=Z_\pi(0)$, see also \eq{eq:propparam}. Note that only LPA'+Y and the fully momentum-dependent truncations PMOM and PVMOM allow a distinction between pion and sigma meson on the level of the momentum dependence of the propagator. Whereas the momentum dependence of the pion propagator is essentially negligible in both fully momentum-dependent truncations, the momentum dependence of the sigma meson propagator is slightly more pronounced. The inclusion of momentum-dependent vertices and in particular the momentum-dependent sigma 4-point function that couples back into the sigma propagator leads to an overall reduction of $\bar Z_\sigma$ along with a slightly less momentum-dependent dressing function in PVMOM compared to PMOM.

The PVMOM scheme includes momentum-dependent 3- and 4-point vertices, whose momentum dependence at the symmetric point momentum configuration is shown in Fig.~\ref{fig:vertices}. Here one should keep in mind that the flow was initialized using momentum-independent 3- and 4-point functions at $\Lambda_{\text{}}=900$~MeV, whereas the full QCD flow initialized at some large perturbative scale will obviously provide momentum-dependent initial conditions. The 3-point vertices and the two-pion-two-sigma as well as the four-pion vertex only show a mild momentum dependence. Interestingly, the momentum dependence of the
sigma 4-point function is comparably strong with a decrease by 37 \% at $p_{\text{sym}}=500$~MeV compared to the value at vanishing external momentum. This is an important observation since the strong momentum dependence occurs in the momentum range below 500~MeV where mesonic fluctuations become quantitatively important in the QCD setting. This suggests that it might have (semi-)quantitative impact also in the full system of QCD correlation functions. However, here one has to keep in mind that the momentum dependence of the sigma meson 4-point function apart from vertices only couples back into the sigma meson propagator, whose impact on the final result is significantly smaller than that of the pion propagator due to its larger mass. 

Relying on the experiences from the Yang-Mills system \cite{Cyrol:2016tym}, the one-dimensional average momentum dependence at the symmetric point is expected to correctly capture at least the qualitative effects of momentum-dependent vertices. The obvious extension along the lines of \cite{Cyrol:2016tym} is the calculation of the momentum dependence of 4-point vertices in the momentum configuration that is required in the tadpole diagrams using momentum-averaged vertices on the right-hand side of the equation. For studies of the full momentum dependence of the 4-point vertex in the symmetric phase of a scalar theory, we refer to \cite{Carrington:2013jta}. The study of critical properties such as critical exponents in the PMOM and PVMOM schemes in 2+1 dimensions at vanishing temperature or in 3+1 dimensions at finite temperature is deferred to future work.

\subsection{Spectral functions}
\label{sec:sfresults}
\begin{figure*}[t]
  \centering \subfloat[Comparing spectral functions obtained in various truncation schemes.\hfill\textcolor{white}{.}]{
\includegraphics[width=0.48\textwidth]{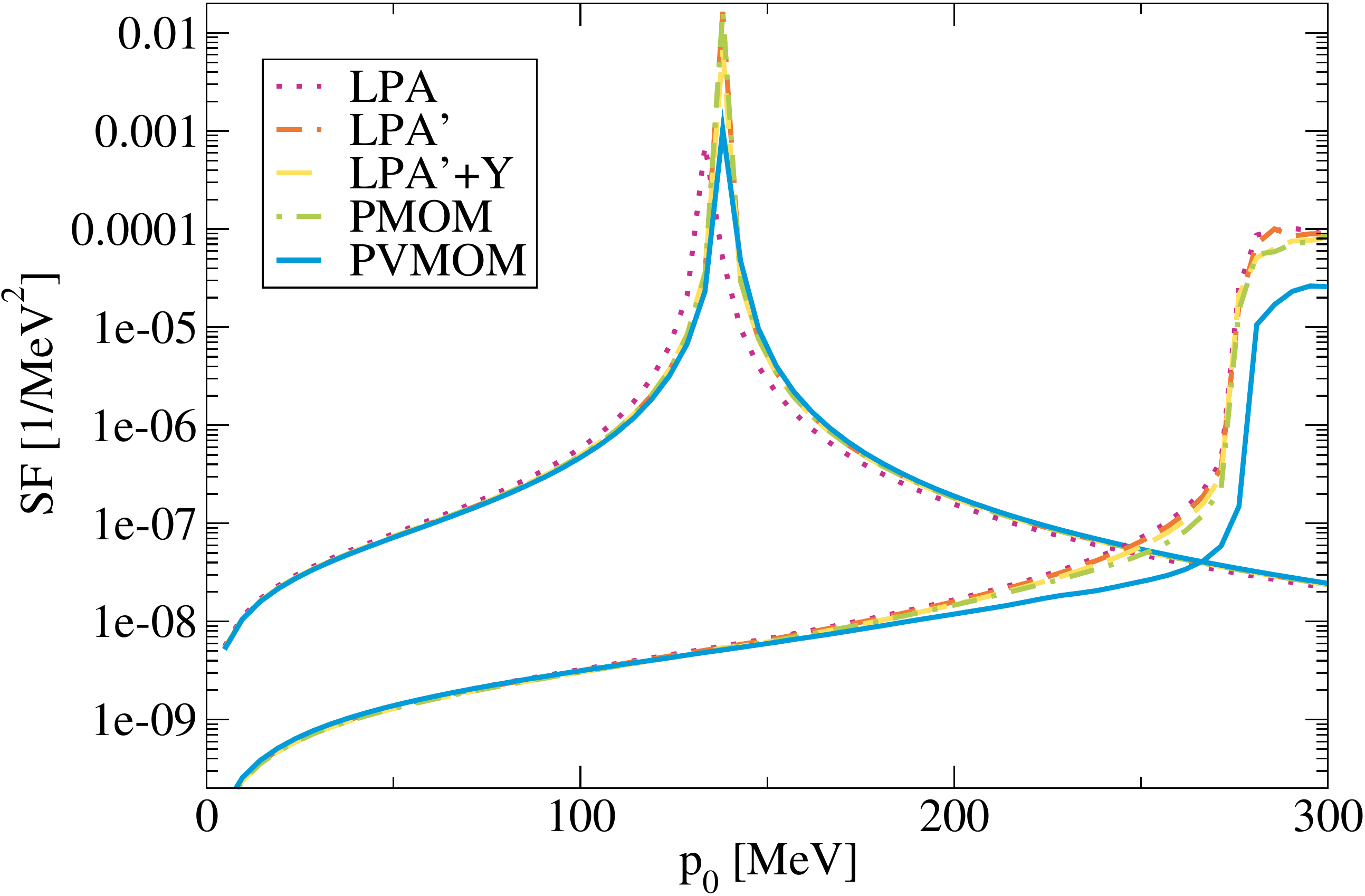}
    \label{fig:comparesf}} \hfill \subfloat[Comparing back coupled pole-corrected propagators (symbols) to pole-corrected propagators evaluated on the simply continued solution (solid line) in the two truncation schemes with the full momentum dependence, PMOM and PVMOM.\hfill\textcolor{white}{.}]{\includegraphics[width=0.47\textwidth]{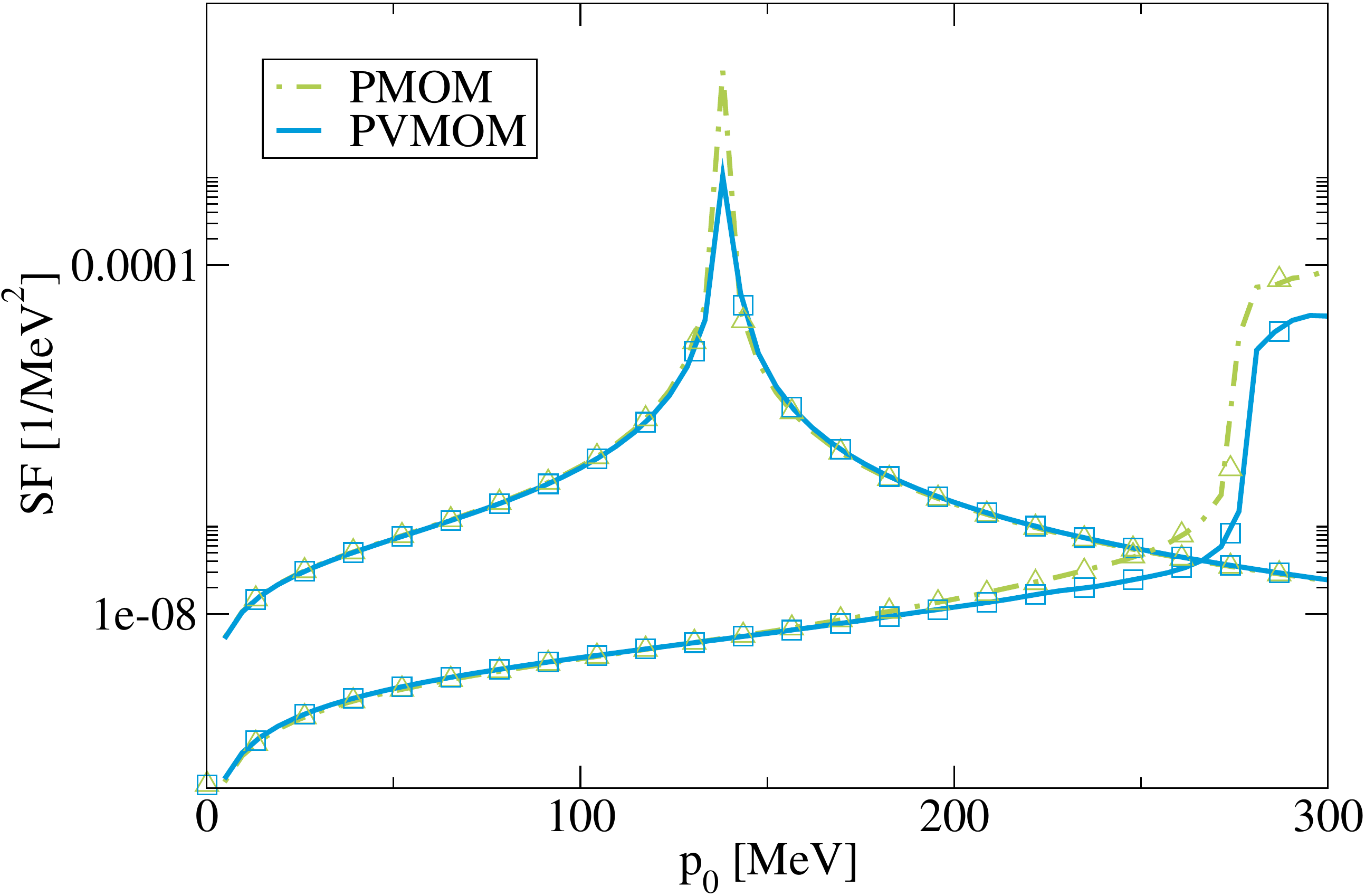}
\label{fig:comparesfcoupled}}
\caption{Pion and sigma meson spectral functions at vanishing spatial external momentum.\hfill\textcolor{white}{.} }\label{fig:sf}
\end{figure*}

In Fig.~\ref{fig:comparesf} we compare the spectral functions obtained in the various truncation schemes. The spectral functions in Fig.~\ref{fig:comparesf} were all obtained by evaluating the pole corrections on a given solution for the simply continued propagators, see also the discussion in Sec.~\ref{sec:spectral}.

The pion spectral function shows a peak that determines by definition the pion pole mass. As demonstrated explicitly in \cite{Pawlowski:2015mia} this spectral function turns into a delta function
at vanishing temperature in the limit $\epsilon\to 0$. The distinct feature in the pion spectral function in the given momentum range is the pion pole mass. Most notably, the LPA spectral function shows a deviation of 3\% to the value of the pion curvature mass that was used to fix the model parameters. The inclusions of fermions leads to an even larger deviation in LPA
as investigated in detail in \cite{Pawlowski:2015mia}. The spectral functions except for the LPA solution show a very similar behavior with nearly coinciding pion pole masses.  

It is worthwhile noting that the spectral functions calculated on a LPA(') solution do not represent
a self-consistent solution as the calculated momentum dependence goes beyond that considered in the Euclidean sector. These solutions should rather be understood as the first iteration
step in an iterative solution procedure analogous to \cite{Helmboldt:2014iya}. This implies in particular that the pole mass calculated from the spectral function
in this first iteration step may deviate from the curvature
mass of the zeroth iteration step which by definition 
coincides with the pole mass in this zeroth step.

\begin{table}[!ht]
  \centering
  \begin{tabular}{l||l||l|l||l|l}
    truncation & $m_{\pi,\text{pole}}$ & $m_{\pi,\text{cur}}$ & $m_{\sigma,\text{cur}}$ & $m^2_{\text{UV}}$ [$\Lambda_{\text{UV}}^2$] & $\lambda_{\text{UV}}$\\
    \hline
    LPA &  134.3 & 138.0 & 384.7 & -0.6367 & 43.64\\
    LPA'&  137.9 & 138.0 & 384.7& -0.7199 & 51.8\\
    LPA'+Y & 137.9 & 138.0 & 384.8 &-0.6423 & 46.51\\
    PMOM &  137.9 & 138.0 & 384.7 & -0.6529 & 47.15\\
    PVMOM & 138.9 & 138.0 & 384.6 & -0.7495 & 55.9
  \end{tabular}
  \caption{Bottom-up comparison of pion pole and curvature mass. For completeness we also indicate the sigma curvature mass as well as the UV parameters of the potential at $\Lambda_{\text{UV}}=900$~MeV.}
  \label{tab:mpi}
\end{table}

In Tab.~\ref{tab:mpi} we compare the pion curvature and pole masses, see \cite{Helmboldt:2014iya} for a comprehensive discussion of different mass definitions, which can be extracted from the analytically continued propagators. Whereas the former is extracted from the zero momentum limit of the propagator, which is most easily accessible in Euclidean approaches, is the latter given by the zero of the propagator at finite Minkowski external momentum, i.e.\
\begin{align}
m_{\pi,\text{cur}}^2=\frac{1}{Z}\Gamma^{(2)}_{\pi\pi}(0,\vec 0)\nonumber\\[2ex]
0=\Gamma^{(2)}_{\pi\pi}(i m_{\pi,\text{pole}},\vec 0)\,.
\end{align}
The curvature mass is unlike the pole mass, which is associated to an inverse temporal screening length, no a direct physical observable. Furthermore there is no fundamental reason that requires the degeneracy of these two mass definitions at vanishing temperature \cite{Helmboldt:2014iya}. Rather, the degree of their agreement
probes the dependence of wavefunction renormalization on Minkowski momenta, which was found to be small for the case of the pion \cite{Helmboldt:2014iya}, which was however just inferred indirectly
by analytically continuing Euclidean data rather than directly calculated. The very good agreement of pion pole and screening masses in the PMOM truncation scheme provides a direct confirmation of this
earlier result. The three different truncation schemes for the momentum dependence of the propagators gradually improve the agreement and the largest improvement already takes place from LPA to LPA'. Interestingly, the deviations in the PVMOM scheme with momentum-dependent vertices increase again slightly, which might still be due to an insufficient momentum resolution of the vertices, but might
also point at a genuine effect of momentum-dependent vertices that has not been considered in earlier studies as there is no fundamental reason for an agreement of curvature and pole masses. Nevertheless
the approximate agreement in all truncation schemes beyond LPA is a good message for studies that fix or measure the pion curvature, which is directly accessible in a purely Euclidean calculation, as a proxy
for the pion pole mass.

Coming back to the discussion of the sigma spectral function in Fig.~\ref{fig:comparesf}, we note that the sigma spectral function shows the 2-pion-decay threshold as characteristic feature in the
momentum range under consideration. Again, in the $\epsilon\to 0$ limit the sigma spectral function at vanishing temperature has to vanish identically below this value.

Interestingly, all truncations with momentum-independent vertices show an approximate agreement of the sigma meson spectral function to a surprisingly high degree of accuracy. The inclusion of momentum-dependent vertices leads to a quantitative
change of the sigma spectral function with a slightly larger value of the two-pion threshold. However, in all cases except in the LPA the two-pion threshold is numerically consistent with twice the value of the
pion pole mass, which is slightly larger in PVMOM compared to the other schemes. The quantitative differences in the case of PVMOM hints at the fact that in order to achieve quantitative accuracy in the investigation of spectral functions, which are inherently momentum-dependent objects, as much (Euclidean but eventually also Minkowski) momentum dependence as possible has to be taken into account.

In Fig.~\ref{fig:comparesfcoupled} we illustrate the backcoupling effects in the pole-corrected propagators on the resulting spectral functions. These effects are indistinguishable on the level of the spectral function itself and stay within the numerical accuracy of the calculation. These findings advocate the use of spectral functions obtained from evaluating pole corrections on the level of the simply continued solution as these calculations also come with a considerably smaller run time compared to the fully back coupled solution where poles have to be traced for the full three-dimensional propagator. However, the
reason for the agreement of both procedures might just lie in the fact that both calculations can only deviate in momentum regions where the complex propagator develops a genuine imaginary part, which in this case just applies to the region beyond the two-pion-threshold in the complex sigma propagator. In addition this applies only to small spatial momenta as the two-pion threshold moves to larger frequencies at finite external momenta as required by Lorentz invariance.

Given these recent advances it will be interesting to follow up on a direct calculation of spectral functions in comparison to
spectral functions inferred indirectly from Euclidean data. The latter are most conveniently provided within FRG studies for example in $\phi^4$ theory \cite{Rose:2016wqz}, critical $O(N)$ models in 2+1 dimensions \cite{Rancon:2014cfa,Rose:2015bma} or (nonrelativistic) Yukawa systems \cite{Schmidt:2011zu}.

\section{Summary and Conclusions}
\label{sec:summary}
In this paper, we presented the first directly calculated self-consistent
spectral functions in the $O(N)$ model which 
include the full (complex) momentum dependence of the propagators as well
as momentum-dependent vertices. This
study illustrates the applicability of the computational framework \cite{Pawlowski:2015mia} 
to more general situations towards quantitative studies of elementary spectral 
functions in QCD.

In the Euclidean sector the results for the truncation with momentum-dependent
propagators represent a nontrivial confirmation of earlier results \cite{Helmboldt:2014iya},
whereas the full truncation goes beyond approaches in the literature, where the
strong momentum dependence of the sigma meson four-point function represents an
interesting observation. In all truncations except in the LPA, the pion pole mass
as the most prominent feature of the pion spectral function in the considered
momentum range, and the pion curvature mass agree approximately. Interestingly, the sigma
spectral function agree approximately in all truncations with momentum-independent vertices, 
whereas the full truncation shows quantitative changes hinting at the importance of 
momentum-dependent truncations in order to achieve
quantitative accuracy in spectral functions.

Apart from direct applications in the $O(N)$ model discussed above and the obvious generalization
to more complex theories with different particle species, which is currently investigated 
for the closely related problem of finite chemical potential in a quark-meson model setting \cite{PSSW}, 
there are two obvious extension of the framework that are both subject to current investigations: The first
one is the extension to finite temperature along the lines of \cite{Pawlowski:2015mia}
which remains to be investigated numerically in a quantitative setting. The second
concerns the extension towards Minkowski momentum dependence in vertex functions
to resolve effects of resonances and additional decay channels that are only captured
in such (Minkowski) momentum-dependent approximation schemes.

\acknowledgments 
\hfill\\
\noindent \emph{Acknowledgments} The author thanks S.~Schlichting for discussions and J.~Pawlowski and N.~Wink for discussions and 
work on related projects. This work was
supported by the DFG under grant no. Str1462/1-1 and 
by the Office of Nuclear Physics in the US Department of
Energy's Office of Science under Contract No. DE-AC02-05CH11231.

\appendix

\section{Vertex expansion in the $O(N)$ model}
\label{app:vertexexp}
In this section, we present details on the vertex expansion in the $O(N)$ model.
The (inverse) propagators are parametrized according to
\begin{align}
\Gamma^{(2)}_{\pi_i\pi_j,k}&=Z_k (\bar Z_{\pi,k}(p) p^2+\bar m_{\pi,k}^2) \delta_{ij}\,,\nonumber\\[2ex]
\Gamma^{(2)}_{\sigma\sigma,k}&=Z_k (\bar Z_{\sigma,k}(p) p^2+\bar m_{\sigma,k}^2)\,,
\label{eq:propparam}
\end{align}
where $\lim_{p\to 0} \bar Z_{\pi,k}(p)p^2=\lim_{p\to 0} \bar Z_{\sigma,k}(p)p^2=0$. $\bar{m}_{\pi,k}^2$ and $\bar{m}_{\sigma,k}^2$ are
renormalized curvature masses that, given \eq{eq:decomposition}, relate to the effective potential
via
\begin{equation}
\bar{m}_{\pi,k}^2=U_k^{(1)}(\rhob)\,,\quad \bar{m}_{\sigma,k}^2=U_k^{(1)}(\rhob)+\,2\rhob\, U_k^{(2)}(\rhob)\,,
\end{equation}
where superscripts are understood as derivatives with respect to $\rhob$. The wavefunction
renormalization $Z_k$ is extracted from the pion propagator evaluated at vanishing momentum. 
The 3-point vertices are parametrized as
\begin{align}
\Gamma^{(3)}_{\pi_i\pi_j\sigma,k}(p_1,p_2)&=Z_k^{3/2} \delta_{ij} \bar Z_{2\pi\sigma,k}(p_1,p_2)\,,\nonumber\\[2ex]
\Gamma^{(3)}_{\sigma\sigma\sigma,k}(p_1,p_2)&=Z_k^{3/2} \bar Z_{3\sigma,k}(p_1,p_2)\,.
\end{align}
Again via \eq{eq:decomposition} these relate to the effective potential in the zero-momentum limit via
\begin{align}
\bar Z_{2\pi\sigma,k}(0,0)&=\sqrt{2\rhob}\, U_k^{(2)}(\rhob)\nonumber\\[2ex]
\bar Z_{3\sigma,k}(0,0)&=\sqrt{2\rhob}\,  \left(3 U_k^{(2)}(\rhob) + 2 \rhob\, U_k^{(3)}(\rhob)\right)
\end{align}
Turning to 4-point vertices, we define
\begin{align}
\Gamma^{(4)}_{\pi_i\pi_j\pi_k\pi_l,k}(p_1,p_2,p_3)&=Z_k^{2} \bar Z_{4\pi,k}(p_1,p_2,p_3)\nonumber\\[2ex]
&\quad \left(\delta_{ij}\delta_{kl}+\delta_{ik}\delta_{jl}+\delta_{il}\delta_{jk}\right)\,,\nonumber\\[2ex]
\Gamma^{(4)}_{\pi_i\pi_j\sigma\sigma,k}(p_1,p_2,p_3)&=Z_k^{2} \bar Z_{2\pi2\sigma,k}(p_1,p_2,p_3) \delta_{ij}\,,\nonumber\\[2ex]
\Gamma^{(4)}_{\sigma\sigma\sigma\sigma,k}(p_1,p_2,p_3)&=Z_k^{2}\bar Z_{4\sigma,k}(p_1,p_2,p_3)\,.
\end{align}
These relate to the effective potential via
\begin{align}
\bar Z_{4\pi,k}(0)&=U_k^{(2)}(\rhob)\,,\nonumber
\end{align}
\begin{align}
\bar Z_{2\pi2\sigma,k}(0)&=U_k^{(2)}(\rhob)+2\rhob U_k^{(3)}(\rhob)\,,\nonumber\\[2ex]
\bar Z_{4\sigma,k}(0)&=3 U_k^{(2)}(\rhob) + 4 \rhob (3 U_k^{(3)}(\rhob) + \rhob U_k^{(4)}(\rhob))\,.
\end{align}
This scheme is extended straightforwardly to higher-order vertex functions
of order 5 and 6, which appear explicitly in the flow equations for the
$4-$ and $3$-point functions, which are approximated momentum independently
using the effective potential.
\section{Flow equations and solution procedure}
\label{app:flows}
In this section, we provide exemplary flow equations used in this work focusing on the flow equations for the (inverse) propagators. The flow equations
for the momentum-dependent contributions are given by
\begin{widetext}
\begin{align}
&\partial_t \Delta\Gamma^{(2)}_{\pi}(p)=\Biggl\{\int_q  \frac{ \dtZr{q} \Zk^3 q^2 \Zpps(-p, -q) \Zpps(p, q)}{( \GammatwopionpR{q}^2 \GammatwosigmapR{p + q}}-\frac{N+1}{2}\int_q\frac{\dtZr{q} \Zk^2 q^2 \Zpppp(p,-p,q)}{\GammatwopionpR{q}^2}\nonumber\\[2ex] 
&+ \int_q\frac{\dtZr{q^2} \Zk^3 (p + q)^2 \Zpps(-p, -q) \Zpps(p, q)}{ \GammatwopionpR{q} \GammatwosigmapR{p + q}^2}-\frac{1}{2} \int_q\frac{ \dtZr{q} \Zk^2 q^2 \Zppss( p, -p, q)}{\GammatwosigmapR{q}^2}\Biggr\}-\Bigl\{p\to 0\Bigr\}\nonumber\\[2ex]
&\partial_t \Delta\Gamma^{(2)}_{\sigma}(p)=\Biggl\{(N-1)\int_q \frac{ \dtZr{q} \Zk^3 q^2 \Zpps(q, -p - q) \Zpps(p + q, -q)}{ \GammatwopionpR{q}^2 \GammatwopionpR{p + q}} - \frac{N- 1}{2}\int_q\frac{ \dtZr{q} \Zk^2 q^2 \Zppss(q, -q, p)}{\GammatwopionpR{q}^2}\nonumber\\[2ex] 
 &+\int_q \frac{ \dtZr{q} \Zk^3 q^2 \Zsss( -p, p + q) \Zsss(p, q)}{
 \GammatwosigmapR{q}^2 \GammatwosigmapR{p + q}} - \frac{1}{2}\int_q\frac{
 \dtZr{q} \Zk^2 q^2 \Zssss(p, -p, q)}{\GammatwosigmapR{q}^2}\Biggr\}-\Bigl\{p\to 0\Bigr\} \label{eq:deltagammaexplicit}
\end{align}
\end{widetext}
where the momentum-independent contribution is provided by the effective potential, see \eq{eq:decomposition}. The flow equation for the effective potential
is simply given by
\begin{equation}
\partial_t U_k=\frac{1}{2}\int_q \left(\frac{(N-1) \dtZr{q} q^2}{\GammatwopionpR{q}}+ \frac{\dtZr{q} q^2}{ \GammatwosigmapR{q}}\right)\,,
\end{equation}
see also \cite{Helmboldt:2014iya}. We refrain from presenting explicit flow equations for the momentum-dependent vertex functions as they represent quite long expressions which do not provide
direct physical insights for the reader. However, given the definitions in App.~\ref{app:vertexexp} these equations can be derived straightforwardly using 
\textit{DoFun} \cite{Huber:2011qr} and \textit{FormTracer} \cite{Cyrol:2016zqb}.

In the PMOM and PVMOM truncations we resolve the full three-dimensional momentum dependence of the propagators, i.e.\ $\Delta\Gamma^{(2)}_{x}(p)\equiv \Delta\Gamma^{(2)}_{x}(p_0^R+\imag p_0^I,|\vec p|)$
for $x\in \{\pi,\sigma\}$. Both objects are discretized on a momentum grid with $10-20$ points in each direction with intermediate values inferred
via spline interpolation. Similarly the vertex equations are solved using a one-dimensional momentum variable at the symmetric point configuration which are due
to the rather mild momentum dependence of the vertex functions already well-approximated using eight grid points. However, we checked in all cases the stability of 
the results upon increasing the number of grid points. In the propagator equations, \eq{eq:deltagammaexplicit}, and the corresponding vertex equations the occurring
momentum-dependent vertex functions are evaluated at average momenta $p^2_{\text{sym}}=\frac{1}{n}(p_1^2+\ldots p_n^2)$. The system of differential equations for expansion coefficients
of the effective potential, the momentum-dependent propagators on a three-dimensional grid and the flow equations for the vertices spanned on a one-dimensional grid
is solved simultaneously and self-consistently within the \textit{frgsolver}, which in turn uses a Runge-Kutta-Cash-Karp adaptive integrator \cite{odeboost}. For 
further details on the numerical procedure the reader is referred to \cite{Cyrol:2016tym}.

\bibliography{../bib_master}

\begin{thebibliography}{60}%
\makeatletter
\providecommand \@ifxundefined [1]{%
 \@ifx{#1\undefined}
}%
\providecommand \@ifnum [1]{%
 \ifnum #1\expandafter \@firstoftwo
 \else \expandafter \@secondoftwo
 \fi
}%
\providecommand \@ifx [1]{%
 \ifx #1\expandafter \@firstoftwo
 \else \expandafter \@secondoftwo
 \fi
}%
\providecommand \natexlab [1]{#1}%
\providecommand \enquote  [1]{``#1''}%
\providecommand \bibnamefont  [1]{#1}%
\providecommand \bibfnamefont [1]{#1}%
\providecommand \citenamefont [1]{#1}%
\providecommand \href@noop [0]{\@secondoftwo}%
\providecommand \href [0]{\begingroup \@sanitize@url \@href}%
\providecommand \@href[1]{\@@startlink{#1}\@@href}%
\providecommand \@@href[1]{\endgroup#1\@@endlink}%
\providecommand \@sanitize@url [0]{\catcode `\\12\catcode `\$12\catcode
  `\&12\catcode `\#12\catcode `\^12\catcode `\_12\catcode `\%12\relax}%
\providecommand \@@startlink[1]{}%
\providecommand \@@endlink[0]{}%
\providecommand \url  [0]{\begingroup\@sanitize@url \@url }%
\providecommand \@url [1]{\endgroup\@href {#1}{\urlprefix }}%
\providecommand \urlprefix  [0]{URL }%
\providecommand \Eprint [0]{\href }%
\providecommand \doibase [0]{http://dx.doi.org/}%
\providecommand \selectlanguage [0]{\@gobble}%
\providecommand \bibinfo  [0]{\@secondoftwo}%
\providecommand \bibfield  [0]{\@secondoftwo}%
\providecommand \translation [1]{[#1]}%
\providecommand \BibitemOpen [0]{}%
\providecommand \bibitemStop [0]{}%
\providecommand \bibitemNoStop [0]{.\EOS\space}%
\providecommand \EOS [0]{\spacefactor3000\relax}%
\providecommand \BibitemShut  [1]{\csname bibitem#1\endcsname}%
\let\auto@bib@innerbib\@empty
\bibitem [{\citenamefont {Strodthoff}\ \emph {et~al.}(2012)\citenamefont
  {Strodthoff}, \citenamefont {Schaefer},\ and\ \citenamefont {von
  Smekal}}]{Strodthoff:2011tz}%
  \BibitemOpen
  \bibfield  {author} {\bibinfo {author} {\bibfnamefont {N.}~\bibnamefont
  {Strodthoff}}, \bibinfo {author} {\bibfnamefont {B.-J.}\ \bibnamefont
  {Schaefer}}, \ and\ \bibinfo {author} {\bibfnamefont {L.}~\bibnamefont {von
  Smekal}},\ }\href {\doibase 10.1103/PhysRevD.85.074007} {\bibfield  {journal}
  {\bibinfo  {journal} {Phys.Rev.}\ }\textbf {\bibinfo {volume} {D85}},\
  \bibinfo {pages} {074007} (\bibinfo {year} {2012})},\ \Eprint
  {http://arxiv.org/abs/1112.5401} {arXiv:1112.5401 [hep-ph]} \BibitemShut
  {NoStop}%
\bibitem [{\citenamefont {Kamikado}\ \emph {et~al.}(2014)\citenamefont
  {Kamikado}, \citenamefont {Strodthoff}, \citenamefont {von Smekal},\ and\
  \citenamefont {Wambach}}]{Kamikado:2013sia}%
  \BibitemOpen
  \bibfield  {author} {\bibinfo {author} {\bibfnamefont {K.}~\bibnamefont
  {Kamikado}}, \bibinfo {author} {\bibfnamefont {N.}~\bibnamefont
  {Strodthoff}}, \bibinfo {author} {\bibfnamefont {L.}~\bibnamefont {von
  Smekal}}, \ and\ \bibinfo {author} {\bibfnamefont {J.}~\bibnamefont
  {Wambach}},\ }\href {\doibase 10.1140/epjc/s10052-014-2806-6} {\bibfield
  {journal} {\bibinfo  {journal} {Eur.Phys.J.}\ }\textbf {\bibinfo {volume}
  {C74}},\ \bibinfo {pages} {2806} (\bibinfo {year} {2014})},\ \Eprint
  {http://arxiv.org/abs/1302.6199} {arXiv:1302.6199 [hep-ph]} \BibitemShut
  {NoStop}%
\bibitem [{\citenamefont {Strauss}\ \emph {et~al.}(2012)\citenamefont
  {Strauss}, \citenamefont {Fischer},\ and\ \citenamefont
  {Kellermann}}]{Strauss:2012dg}%
  \BibitemOpen
  \bibfield  {author} {\bibinfo {author} {\bibfnamefont {S.}~\bibnamefont
  {Strauss}}, \bibinfo {author} {\bibfnamefont {C.~S.}\ \bibnamefont
  {Fischer}}, \ and\ \bibinfo {author} {\bibfnamefont {C.}~\bibnamefont
  {Kellermann}},\ }\href {\doibase 10.1103/PhysRevLett.109.252001} {\bibfield
  {journal} {\bibinfo  {journal} {Phys.Rev.Lett.}\ }\textbf {\bibinfo {volume}
  {109}},\ \bibinfo {pages} {252001} (\bibinfo {year} {2012})},\ \Eprint
  {http://arxiv.org/abs/1208.6239} {arXiv:1208.6239 [hep-ph]} \BibitemShut
  {NoStop}%
\bibitem [{\citenamefont {Pawlowski}\ and\ \citenamefont
  {Strodthoff}(2015)}]{Pawlowski:2015mia}%
  \BibitemOpen
  \bibfield  {author} {\bibinfo {author} {\bibfnamefont {J.~M.}\ \bibnamefont
  {Pawlowski}}\ and\ \bibinfo {author} {\bibfnamefont {N.}~\bibnamefont
  {Strodthoff}},\ }\href {\doibase 10.1103/PhysRevD.92.094009} {\bibfield
  {journal} {\bibinfo  {journal} {Phys. Rev.}\ }\textbf {\bibinfo {volume}
  {D92}},\ \bibinfo {pages} {094009} (\bibinfo {year} {2015})},\ \Eprint
  {http://arxiv.org/abs/1508.01160} {arXiv:1508.01160 [hep-ph]} \BibitemShut
  {NoStop}%
\bibitem [{\citenamefont {Canet}\ and\ \citenamefont
  {Chat{\'e}}(2007)}]{Canet:2006xu}%
  \BibitemOpen
  \bibfield  {author} {\bibinfo {author} {\bibfnamefont {L.}~\bibnamefont
  {Canet}}\ and\ \bibinfo {author} {\bibfnamefont {H.}~\bibnamefont
  {Chat{\'e}}},\ }\href {\doibase 10.1088/1751-8113/40/9/002} {\bibfield
  {journal} {\bibinfo  {journal} {J. Phys.}\ }\textbf {\bibinfo {volume}
  {40}},\ \bibinfo {pages} {1937} (\bibinfo {year} {2007})},\ \Eprint
  {http://arxiv.org/abs/cond-mat/0610468} {arXiv:cond-mat/0610468
  [cond-mat.stat-mech]} \BibitemShut {NoStop}%
\bibitem [{\citenamefont {Mesterh\'{a}zy}\ \emph {et~al.}(2013)\citenamefont
  {Mesterh\'{a}zy}, \citenamefont {Stockemer}, \citenamefont {Palhares},\ and\
  \citenamefont {Berges}}]{Mesterhazy:2013naa}%
  \BibitemOpen
  \bibfield  {author} {\bibinfo {author} {\bibfnamefont {D.}~\bibnamefont
  {Mesterh\'{a}zy}}, \bibinfo {author} {\bibfnamefont {J.~H.}\ \bibnamefont
  {Stockemer}}, \bibinfo {author} {\bibfnamefont {L.~F.}\ \bibnamefont
  {Palhares}}, \ and\ \bibinfo {author} {\bibfnamefont {J.}~\bibnamefont
  {Berges}},\ }\href {\doibase 10.1103/PhysRevB.88.174301} {\bibfield
  {journal} {\bibinfo  {journal} {Phys.Rev.}\ }\textbf {\bibinfo {volume}
  {B88}},\ \bibinfo {pages} {174301} (\bibinfo {year} {2013})},\ \Eprint
  {http://arxiv.org/abs/1307.1700} {arXiv:1307.1700} \BibitemShut {NoStop}%
\bibitem [{\citenamefont {Mesterh\'{a}zy}\ \emph {et~al.}(2015)\citenamefont
  {Mesterh\'{a}zy}, \citenamefont {Stockemer},\ and\ \citenamefont
  {Tanizaki}}]{Mesterhazy:2015uja}%
  \BibitemOpen
  \bibfield  {author} {\bibinfo {author} {\bibfnamefont {D.}~\bibnamefont
  {Mesterh\'{a}zy}}, \bibinfo {author} {\bibfnamefont {J.~H.}\ \bibnamefont
  {Stockemer}}, \ and\ \bibinfo {author} {\bibfnamefont {Y.}~\bibnamefont
  {Tanizaki}},\ }\href {\doibase 10.1103/PhysRevD.92.076001} {\bibfield
  {journal} {\bibinfo  {journal} {Phys. Rev.}\ }\textbf {\bibinfo {volume}
  {D92}},\ \bibinfo {pages} {076001} (\bibinfo {year} {2015})},\ \Eprint
  {http://arxiv.org/abs/1504.07268} {arXiv:1504.07268 [hep-ph]} \BibitemShut
  {NoStop}%
\bibitem [{\citenamefont {Kyung}\ \emph {et~al.}(1998)\citenamefont {Kyung},
  \citenamefont {Klepfish},\ and\ \citenamefont {Kornilovitch}}]{Kyung1998}%
  \BibitemOpen
  \bibfield  {author} {\bibinfo {author} {\bibfnamefont {B.}~\bibnamefont
  {Kyung}}, \bibinfo {author} {\bibfnamefont {E.~G.}\ \bibnamefont {Klepfish}},
  \ and\ \bibinfo {author} {\bibfnamefont {P.~E.}\ \bibnamefont
  {Kornilovitch}},\ }\href {\doibase 10.1103/PhysRevLett.80.3109} {\bibfield
  {journal} {\bibinfo  {journal} {Phys. Rev. Lett.}\ }\textbf {\bibinfo
  {volume} {80}},\ \bibinfo {pages} {3109} (\bibinfo {year} {1998})},\ \Eprint
  {http://arxiv.org/abs/cond-mat/9802239v1} {cond-mat/9802239v1} \BibitemShut
  {NoStop}%
\bibitem [{\citenamefont {Rohe}\ and\ \citenamefont
  {Metzner}(2001)}]{Rohe2000}%
  \BibitemOpen
  \bibfield  {author} {\bibinfo {author} {\bibfnamefont {D.}~\bibnamefont
  {Rohe}}\ and\ \bibinfo {author} {\bibfnamefont {W.}~\bibnamefont {Metzner}},\
  }\href {\doibase 10.1103/PhysRevB.63.224509} {\bibfield  {journal} {\bibinfo
  {journal} {Phys. Rev.}\ }\textbf {\bibinfo {volume} {B63}},\ \bibinfo {pages}
  {224509} (\bibinfo {year} {2001})},\ \Eprint
  {http://arxiv.org/abs/cond-mat/0011500v2} {cond-mat/0011500v2} \BibitemShut
  {NoStop}%
\bibitem [{\citenamefont {Tripolt}\ \emph
  {et~al.}(2014{\natexlab{a}})\citenamefont {Tripolt}, \citenamefont
  {Strodthoff}, \citenamefont {von Smekal},\ and\ \citenamefont
  {Wambach}}]{Tripolt:2013jra}%
  \BibitemOpen
  \bibfield  {author} {\bibinfo {author} {\bibfnamefont {R.-A.}\ \bibnamefont
  {Tripolt}}, \bibinfo {author} {\bibfnamefont {N.}~\bibnamefont {Strodthoff}},
  \bibinfo {author} {\bibfnamefont {L.}~\bibnamefont {von Smekal}}, \ and\
  \bibinfo {author} {\bibfnamefont {J.}~\bibnamefont {Wambach}},\ }\href
  {\doibase 10.1103/PhysRevD.89.034010} {\bibfield  {journal} {\bibinfo
  {journal} {Phys.Rev.}\ }\textbf {\bibinfo {volume} {D89}},\ \bibinfo {pages}
  {034010} (\bibinfo {year} {2014}{\natexlab{a}})},\ \Eprint
  {http://arxiv.org/abs/1311.0630} {arXiv:1311.0630 [hep-ph]} \BibitemShut
  {NoStop}%
\bibitem [{\citenamefont {Tripolt}\ \emph
  {et~al.}(2014{\natexlab{b}})\citenamefont {Tripolt}, \citenamefont {von
  Smekal},\ and\ \citenamefont {Wambach}}]{Tripolt:2014wra}%
  \BibitemOpen
  \bibfield  {author} {\bibinfo {author} {\bibfnamefont {R.-A.}\ \bibnamefont
  {Tripolt}}, \bibinfo {author} {\bibfnamefont {L.}~\bibnamefont {von Smekal}},
  \ and\ \bibinfo {author} {\bibfnamefont {J.}~\bibnamefont {Wambach}},\ }\href
  {\doibase 10.1103/PhysRevD.90.074031} {\bibfield  {journal} {\bibinfo
  {journal} {Phys.Rev.}\ }\textbf {\bibinfo {volume} {D90}},\ \bibinfo {pages}
  {074031} (\bibinfo {year} {2014}{\natexlab{b}})},\ \Eprint
  {http://arxiv.org/abs/1408.3512} {arXiv:1408.3512 [hep-ph]} \BibitemShut
  {NoStop}%
\bibitem [{\citenamefont {Tripolt}\ \emph {et~al.}(2017)\citenamefont
  {Tripolt}, \citenamefont {von Smekal},\ and\ \citenamefont
  {Wambach}}]{Tripolt:2016cey}%
  \BibitemOpen
  \bibfield  {author} {\bibinfo {author} {\bibfnamefont {R.-A.}\ \bibnamefont
  {Tripolt}}, \bibinfo {author} {\bibfnamefont {L.}~\bibnamefont {von Smekal}},
  \ and\ \bibinfo {author} {\bibfnamefont {J.}~\bibnamefont {Wambach}},\ }\href
  {\doibase 10.1142/S0218301317400286} {\bibfield  {journal} {\bibinfo
  {journal} {Int. J. Mod. Phys.}\ }\textbf {\bibinfo {volume} {E26}},\ \bibinfo
  {pages} {1740028} (\bibinfo {year} {2017})},\ \Eprint
  {http://arxiv.org/abs/1605.00771} {arXiv:1605.00771 [hep-ph]} \BibitemShut
  {NoStop}%
\bibitem [{\citenamefont {Yokota}\ \emph {et~al.}(2016)\citenamefont {Yokota},
  \citenamefont {Kunihiro},\ and\ \citenamefont {Morita}}]{Yokota:2016tip}%
  \BibitemOpen
  \bibfield  {author} {\bibinfo {author} {\bibfnamefont {T.}~\bibnamefont
  {Yokota}}, \bibinfo {author} {\bibfnamefont {T.}~\bibnamefont {Kunihiro}}, \
  and\ \bibinfo {author} {\bibfnamefont {K.}~\bibnamefont {Morita}},\ }\href
  {\doibase 10.1093/ptep/ptw062} {\bibfield  {journal} {\bibinfo  {journal}
  {PTEP}\ }\textbf {\bibinfo {volume} {2016}},\ \bibinfo {pages} {073D01}
  (\bibinfo {year} {2016})},\ \Eprint {http://arxiv.org/abs/1603.02147}
  {arXiv:1603.02147 [hep-ph]} \BibitemShut {NoStop}%
\bibitem [{\citenamefont {Kamikado}\ \emph {et~al.}(2017)\citenamefont
  {Kamikado}, \citenamefont {Kanazawa},\ and\ \citenamefont
  {Uchino}}]{Kamikado:2016chk}%
  \BibitemOpen
  \bibfield  {author} {\bibinfo {author} {\bibfnamefont {K.}~\bibnamefont
  {Kamikado}}, \bibinfo {author} {\bibfnamefont {T.}~\bibnamefont {Kanazawa}},
  \ and\ \bibinfo {author} {\bibfnamefont {S.}~\bibnamefont {Uchino}},\ }\href
  {\doibase 10.1103/PhysRevA.95.013612} {\bibfield  {journal} {\bibinfo
  {journal} {Phys. Rev.}\ }\textbf {\bibinfo {volume} {A95}},\ \bibinfo {pages}
  {013612} (\bibinfo {year} {2017})},\ \Eprint
  {http://arxiv.org/abs/1606.03721} {arXiv:1606.03721 [cond-mat.quant-gas]}
  \BibitemShut {NoStop}%
\bibitem [{\citenamefont {Jung}\ \emph {et~al.}(2017)\citenamefont {Jung},
  \citenamefont {Rennecke}, \citenamefont {Tripolt}, \citenamefont {von
  Smekal},\ and\ \citenamefont {Wambach}}]{Jung:2016yxl}%
  \BibitemOpen
  \bibfield  {author} {\bibinfo {author} {\bibfnamefont {C.}~\bibnamefont
  {Jung}}, \bibinfo {author} {\bibfnamefont {F.}~\bibnamefont {Rennecke}},
  \bibinfo {author} {\bibfnamefont {R.-A.}\ \bibnamefont {Tripolt}}, \bibinfo
  {author} {\bibfnamefont {L.}~\bibnamefont {von Smekal}}, \ and\ \bibinfo
  {author} {\bibfnamefont {J.}~\bibnamefont {Wambach}},\ }\href {\doibase
  10.1103/PhysRevD.95.036020} {\bibfield  {journal} {\bibinfo  {journal} {Phys.
  Rev.}\ }\textbf {\bibinfo {volume} {D95}},\ \bibinfo {pages} {036020}
  (\bibinfo {year} {2017})},\ \Eprint {http://arxiv.org/abs/1610.08754}
  {arXiv:1610.08754 [hep-ph]} \BibitemShut {NoStop}%
\bibitem [{fQC()}]{fQCD:2016-10}%
  \BibitemOpen
  \href@noop {} {}\bibinfo {note} {{fQCD Collaboration, J. Braun, L. Corell,
  A.~K. Cyrol, W.-j. Fu, M. Leonhardt, M. Mitter, J.~M. Pawlowski, M. Pospiech,
  F. Rennecke, N. Strodthoff, N. Wink}}\BibitemShut {NoStop}%
\bibitem [{\citenamefont {Mitter}\ \emph {et~al.}(2015)\citenamefont {Mitter},
  \citenamefont {Pawlowski},\ and\ \citenamefont
  {Strodthoff}}]{Mitter:2014wpa}%
  \BibitemOpen
  \bibfield  {author} {\bibinfo {author} {\bibfnamefont {M.}~\bibnamefont
  {Mitter}}, \bibinfo {author} {\bibfnamefont {J.~M.}\ \bibnamefont
  {Pawlowski}}, \ and\ \bibinfo {author} {\bibfnamefont {N.}~\bibnamefont
  {Strodthoff}},\ }\href {\doibase 10.1103/PhysRevD.91.054035} {\bibfield
  {journal} {\bibinfo  {journal} {Phys.Rev.}\ }\textbf {\bibinfo {volume}
  {D91}},\ \bibinfo {pages} {054035} (\bibinfo {year} {2015})},\ \Eprint
  {http://arxiv.org/abs/1411.7978} {arXiv:1411.7978 [hep-ph]} \BibitemShut
  {NoStop}%
\bibitem [{\citenamefont {Braun}\ \emph {et~al.}(2016)\citenamefont {Braun},
  \citenamefont {Fister}, \citenamefont {Pawlowski},\ and\ \citenamefont
  {Rennecke}}]{Braun:2014ata}%
  \BibitemOpen
  \bibfield  {author} {\bibinfo {author} {\bibfnamefont {J.}~\bibnamefont
  {Braun}}, \bibinfo {author} {\bibfnamefont {L.}~\bibnamefont {Fister}},
  \bibinfo {author} {\bibfnamefont {J.~M.}\ \bibnamefont {Pawlowski}}, \ and\
  \bibinfo {author} {\bibfnamefont {F.}~\bibnamefont {Rennecke}},\ }\href
  {\doibase 10.1103/PhysRevD.94.034016} {\bibfield  {journal} {\bibinfo
  {journal} {Phys. Rev.}\ }\textbf {\bibinfo {volume} {D94}},\ \bibinfo {pages}
  {034016} (\bibinfo {year} {2016})},\ \Eprint {http://arxiv.org/abs/1412.1045}
  {arXiv:1412.1045 [hep-ph]} \BibitemShut {NoStop}%
\bibitem [{\citenamefont {Cyrol}\ \emph
  {et~al.}(2016{\natexlab{a}})\citenamefont {Cyrol}, \citenamefont {Fister},
  \citenamefont {Mitter}, \citenamefont {Pawlowski},\ and\ \citenamefont
  {Strodthoff}}]{Cyrol:2016tym}%
  \BibitemOpen
  \bibfield  {author} {\bibinfo {author} {\bibfnamefont {A.~K.}\ \bibnamefont
  {Cyrol}}, \bibinfo {author} {\bibfnamefont {L.}~\bibnamefont {Fister}},
  \bibinfo {author} {\bibfnamefont {M.}~\bibnamefont {Mitter}}, \bibinfo
  {author} {\bibfnamefont {J.~M.}\ \bibnamefont {Pawlowski}}, \ and\ \bibinfo
  {author} {\bibfnamefont {N.}~\bibnamefont {Strodthoff}},\ }\href {\doibase
  10.1103/PhysRevD.94.054005} {\bibfield  {journal} {\bibinfo  {journal} {Phys.
  Rev.}\ }\textbf {\bibinfo {volume} {D94}},\ \bibinfo {pages} {054005}
  (\bibinfo {year} {2016}{\natexlab{a}})},\ \Eprint
  {http://arxiv.org/abs/1605.01856} {arXiv:1605.01856 [hep-ph]} \BibitemShut
  {NoStop}%
\bibitem [{\citenamefont {Haas}\ \emph {et~al.}(2014)\citenamefont {Haas},
  \citenamefont {Fister},\ and\ \citenamefont {Pawlowski}}]{Haas:2013hpa}%
  \BibitemOpen
  \bibfield  {author} {\bibinfo {author} {\bibfnamefont {M.}~\bibnamefont
  {Haas}}, \bibinfo {author} {\bibfnamefont {L.}~\bibnamefont {Fister}}, \ and\
  \bibinfo {author} {\bibfnamefont {J.~M.}\ \bibnamefont {Pawlowski}},\ }\href
  {\doibase 10.1103/PhysRevD.90.091501} {\bibfield  {journal} {\bibinfo
  {journal} {Phys.Rev.}\ }\textbf {\bibinfo {volume} {D90}},\ \bibinfo {pages}
  {091501} (\bibinfo {year} {2014})},\ \Eprint {http://arxiv.org/abs/1308.4960}
  {arXiv:1308.4960 [hep-ph]} \BibitemShut {NoStop}%
\bibitem [{\citenamefont {Christiansen}\ \emph {et~al.}(2015)\citenamefont
  {Christiansen}, \citenamefont {Haas}, \citenamefont {Pawlowski},\ and\
  \citenamefont {Strodthoff}}]{Christiansen:2014ypa}%
  \BibitemOpen
  \bibfield  {author} {\bibinfo {author} {\bibfnamefont {N.}~\bibnamefont
  {Christiansen}}, \bibinfo {author} {\bibfnamefont {M.}~\bibnamefont {Haas}},
  \bibinfo {author} {\bibfnamefont {J.~M.}\ \bibnamefont {Pawlowski}}, \ and\
  \bibinfo {author} {\bibfnamefont {N.}~\bibnamefont {Strodthoff}},\ }\href
  {\doibase 10.1103/PhysRevLett.115.112002} {\bibfield  {journal} {\bibinfo
  {journal} {Phys. Rev. Lett.}\ }\textbf {\bibinfo {volume} {115}},\ \bibinfo
  {pages} {112002} (\bibinfo {year} {2015})},\ \Eprint
  {http://arxiv.org/abs/1411.7986} {arXiv:1411.7986 [hep-ph]} \BibitemShut
  {NoStop}%
\bibitem [{\citenamefont {Wetterich}(1993{\natexlab{a}})}]{Wetterich:1992yh}%
  \BibitemOpen
  \bibfield  {author} {\bibinfo {author} {\bibfnamefont {C.}~\bibnamefont
  {Wetterich}},\ }\href {\doibase 10.1016/0370-2693(93)90726-X} {\bibfield
  {journal} {\bibinfo  {journal} {Phys.Lett.}\ }\textbf {\bibinfo {volume}
  {B301}},\ \bibinfo {pages} {90} (\bibinfo {year}
  {1993}{\natexlab{a}})}\BibitemShut {NoStop}%
\bibitem [{\citenamefont {Tetradis}\ and\ \citenamefont
  {Wetterich}(1994)}]{Tetradis:1993ts}%
  \BibitemOpen
  \bibfield  {author} {\bibinfo {author} {\bibfnamefont {N.}~\bibnamefont
  {Tetradis}}\ and\ \bibinfo {author} {\bibfnamefont {C.}~\bibnamefont
  {Wetterich}},\ }\href {\doibase 10.1016/0550-3213(94)90446-4} {\bibfield
  {journal} {\bibinfo  {journal} {Nucl.Phys.}\ }\textbf {\bibinfo {volume}
  {B422}},\ \bibinfo {pages} {541} (\bibinfo {year} {1994})},\ \Eprint
  {http://arxiv.org/abs/hep-ph/9308214} {arXiv:hep-ph/9308214 [hep-ph]}
  \BibitemShut {NoStop}%
\bibitem [{\citenamefont {Morris}\ and\ \citenamefont
  {Turner}(1998)}]{Morris:1997xj}%
  \BibitemOpen
  \bibfield  {author} {\bibinfo {author} {\bibfnamefont {T.~R.}\ \bibnamefont
  {Morris}}\ and\ \bibinfo {author} {\bibfnamefont {M.~D.}\ \bibnamefont
  {Turner}},\ }\href {\doibase 10.1016/S0550-3213(97)00640-8} {\bibfield
  {journal} {\bibinfo  {journal} {Nucl. Phys.}\ }\textbf {\bibinfo {volume}
  {B509}},\ \bibinfo {pages} {637} (\bibinfo {year} {1998})},\ \Eprint
  {http://arxiv.org/abs/hep-th/9704202} {arXiv:hep-th/9704202 [hep-th]}
  \BibitemShut {NoStop}%
\bibitem [{\citenamefont {Litim}(2002)}]{Litim:2002cf}%
  \BibitemOpen
  \bibfield  {author} {\bibinfo {author} {\bibfnamefont {D.~F.}\ \bibnamefont
  {Litim}},\ }\href {\doibase 10.1016/S0550-3213(02)00186-4} {\bibfield
  {journal} {\bibinfo  {journal} {Nucl.Phys.}\ }\textbf {\bibinfo {volume}
  {B631}},\ \bibinfo {pages} {128} (\bibinfo {year} {2002})},\ \Eprint
  {http://arxiv.org/abs/hep-th/0203006} {arXiv:hep-th/0203006 [hep-th]}
  \BibitemShut {NoStop}%
\bibitem [{\citenamefont {Blaizot}\ \emph {et~al.}(2006)\citenamefont
  {Blaizot}, \citenamefont {Mend{\'e}z-Galain},\ and\ \citenamefont
  {Wschebor}}]{Blaizot:2006vr}%
  \BibitemOpen
  \bibfield  {author} {\bibinfo {author} {\bibfnamefont {J.-P.}\ \bibnamefont
  {Blaizot}}, \bibinfo {author} {\bibfnamefont {R.}~\bibnamefont
  {Mend{\'e}z-Galain}}, \ and\ \bibinfo {author} {\bibfnamefont
  {N.}~\bibnamefont {Wschebor}},\ }\href {\doibase 10.1103/PhysRevE.74.051117}
  {\bibfield  {journal} {\bibinfo  {journal} {Phys.Rev.}\ }\textbf {\bibinfo
  {volume} {E74}},\ \bibinfo {pages} {051117} (\bibinfo {year} {2006})},\
  \Eprint {http://arxiv.org/abs/hep-th/0603163} {arXiv:hep-th/0603163 [hep-th]}
  \BibitemShut {NoStop}%
\bibitem [{\citenamefont {Blaizot}\ \emph {et~al.}(2007)\citenamefont
  {Blaizot}, \citenamefont {Ipp}, \citenamefont {Mend{\'e}z-Galain},\ and\
  \citenamefont {Wschebor}}]{Blaizot:2006rj}%
  \BibitemOpen
  \bibfield  {author} {\bibinfo {author} {\bibfnamefont {J.-P.}\ \bibnamefont
  {Blaizot}}, \bibinfo {author} {\bibfnamefont {A.}~\bibnamefont {Ipp}},
  \bibinfo {author} {\bibfnamefont {R.}~\bibnamefont {Mend{\'e}z-Galain}}, \
  and\ \bibinfo {author} {\bibfnamefont {N.}~\bibnamefont {Wschebor}},\ }\href
  {\doibase 10.1016/j.nuclphysa.2006.11.139} {\bibfield  {journal} {\bibinfo
  {journal} {Nucl.Phys.}\ }\textbf {\bibinfo {volume} {A784}},\ \bibinfo
  {pages} {376} (\bibinfo {year} {2007})},\ \Eprint
  {http://arxiv.org/abs/hep-ph/0610004} {arXiv:hep-ph/0610004 [hep-ph]}
  \BibitemShut {NoStop}%
\bibitem [{\citenamefont {Benitez}\ \emph {et~al.}(2012)\citenamefont
  {Benitez}, \citenamefont {Blaizot}, \citenamefont {Chat{\'e}}, \citenamefont
  {Delamotte}, \citenamefont {Mend{\'e}z-Galain} \emph
  {et~al.}}]{Benitez:2011xx}%
  \BibitemOpen
  \bibfield  {author} {\bibinfo {author} {\bibfnamefont {F.}~\bibnamefont
  {Benitez}}, \bibinfo {author} {\bibfnamefont {J.-P.}\ \bibnamefont
  {Blaizot}}, \bibinfo {author} {\bibfnamefont {H.}~\bibnamefont {Chat{\'e}}},
  \bibinfo {author} {\bibfnamefont {B.}~\bibnamefont {Delamotte}}, \bibinfo
  {author} {\bibfnamefont {R.}~\bibnamefont {Mend{\'e}z-Galain}},  \emph
  {et~al.},\ }\href {\doibase 10.1103/PhysRevE.85.026707} {\bibfield  {journal}
  {\bibinfo  {journal} {Phys.Rev.}\ }\textbf {\bibinfo {volume} {E85}},\
  \bibinfo {pages} {026707} (\bibinfo {year} {2012})},\ \Eprint
  {http://arxiv.org/abs/1110.2665} {arXiv:1110.2665 [cond-mat.stat-mech]}
  \BibitemShut {NoStop}%
\bibitem [{\citenamefont {Zappala}(2012)}]{Zappala:2012wh}%
  \BibitemOpen
  \bibfield  {author} {\bibinfo {author} {\bibfnamefont {D.}~\bibnamefont
  {Zappala}},\ }\href {\doibase 10.1103/PhysRevD.86.125003} {\bibfield
  {journal} {\bibinfo  {journal} {Phys. Rev.}\ }\textbf {\bibinfo {volume}
  {D86}},\ \bibinfo {pages} {125003} (\bibinfo {year} {2012})},\ \Eprint
  {http://arxiv.org/abs/1206.2480} {arXiv:1206.2480 [hep-th]} \BibitemShut
  {NoStop}%
\bibitem [{\citenamefont {Nagy}(2012)}]{Nagy:2012np}%
  \BibitemOpen
  \bibfield  {author} {\bibinfo {author} {\bibfnamefont {S.}~\bibnamefont
  {Nagy}},\ }\href {\doibase 10.1103/PhysRevD.86.085020} {\bibfield  {journal}
  {\bibinfo  {journal} {Phys. Rev.}\ }\textbf {\bibinfo {volume} {D86}},\
  \bibinfo {pages} {085020} (\bibinfo {year} {2012})},\ \Eprint
  {http://arxiv.org/abs/1201.1625} {arXiv:1201.1625 [hep-th]} \BibitemShut
  {NoStop}%
\bibitem [{\citenamefont {Ran{\c c}on}\ and\ \citenamefont
  {Dupuis}(2014)}]{Rancon:2014cfa}%
  \BibitemOpen
  \bibfield  {author} {\bibinfo {author} {\bibfnamefont {A.}~\bibnamefont
  {Ran{\c c}on}}\ and\ \bibinfo {author} {\bibfnamefont {N.}~\bibnamefont
  {Dupuis}},\ }\href {\doibase 10.1103/PhysRevB.89.180501} {\bibfield
  {journal} {\bibinfo  {journal} {Phys. Rev.}\ }\textbf {\bibinfo {volume}
  {B89}},\ \bibinfo {pages} {180501} (\bibinfo {year} {2014})}\BibitemShut
  {NoStop}%
\bibitem [{\citenamefont {Rose}\ \emph {et~al.}(2015)\citenamefont {Rose},
  \citenamefont {L{\'e}onard},\ and\ \citenamefont {Dupuis}}]{Rose:2015bma}%
  \BibitemOpen
  \bibfield  {author} {\bibinfo {author} {\bibfnamefont {F.}~\bibnamefont
  {Rose}}, \bibinfo {author} {\bibfnamefont {F.}~\bibnamefont {L{\'e}onard}}, \
  and\ \bibinfo {author} {\bibfnamefont {N.}~\bibnamefont {Dupuis}},\ }\href
  {\doibase 10.1103/PhysRevB.91.224501} {\bibfield  {journal} {\bibinfo
  {journal} {Phys. Rev.}\ }\textbf {\bibinfo {volume} {B91}},\ \bibinfo {pages}
  {224501} (\bibinfo {year} {2015})},\ \Eprint
  {http://arxiv.org/abs/1503.08688} {arXiv:1503.08688 [cond-mat.quant-gas]}
  \BibitemShut {NoStop}%
\bibitem [{\citenamefont {Pel{\'a}ez}\ and\ \citenamefont
  {Wschebor}(2016)}]{Pelaez:2015nsa}%
  \BibitemOpen
  \bibfield  {author} {\bibinfo {author} {\bibfnamefont {M.}~\bibnamefont
  {Pel{\'a}ez}}\ and\ \bibinfo {author} {\bibfnamefont {N.}~\bibnamefont
  {Wschebor}},\ }\href {\doibase 10.1103/PhysRevE.94.042136} {\bibfield
  {journal} {\bibinfo  {journal} {Phys. Rev.}\ }\textbf {\bibinfo {volume}
  {E94}},\ \bibinfo {pages} {042136} (\bibinfo {year} {2016})},\ \Eprint
  {http://arxiv.org/abs/1510.05709} {arXiv:1510.05709 [cond-mat.stat-mech]}
  \BibitemShut {NoStop}%
\bibitem [{\citenamefont {Lenaghan}\ and\ \citenamefont
  {Rischke}(2000)}]{Lenaghan:1999si}%
  \BibitemOpen
  \bibfield  {author} {\bibinfo {author} {\bibfnamefont {J.~T.}\ \bibnamefont
  {Lenaghan}}\ and\ \bibinfo {author} {\bibfnamefont {D.~H.}\ \bibnamefont
  {Rischke}},\ }\href {\doibase 10.1088/0954-3899/26/4/309} {\bibfield
  {journal} {\bibinfo  {journal} {J. Phys.}\ }\textbf {\bibinfo {volume}
  {G26}},\ \bibinfo {pages} {431} (\bibinfo {year} {2000})},\ \Eprint
  {http://arxiv.org/abs/nucl-th/9901049} {arXiv:nucl-th/9901049 [nucl-th]}
  \BibitemShut {NoStop}%
\bibitem [{\citenamefont {Nemoto}\ \emph {et~al.}(2000)\citenamefont {Nemoto},
  \citenamefont {Naito},\ and\ \citenamefont {Oka}}]{Nemoto:1999qf}%
  \BibitemOpen
  \bibfield  {author} {\bibinfo {author} {\bibfnamefont {Y.}~\bibnamefont
  {Nemoto}}, \bibinfo {author} {\bibfnamefont {K.}~\bibnamefont {Naito}}, \
  and\ \bibinfo {author} {\bibfnamefont {M.}~\bibnamefont {Oka}},\ }\href
  {\doibase 10.1007/s100500070042} {\bibfield  {journal} {\bibinfo  {journal}
  {Eur.Phys.J.}\ }\textbf {\bibinfo {volume} {A9}},\ \bibinfo {pages} {245}
  (\bibinfo {year} {2000})},\ \Eprint {http://arxiv.org/abs/hep-ph/9911431}
  {arXiv:hep-ph/9911431 [hep-ph]} \BibitemShut {NoStop}%
\bibitem [{\citenamefont {Roder}\ \emph {et~al.}(2006)\citenamefont {Roder},
  \citenamefont {Ruppert},\ and\ \citenamefont {Rischke}}]{Roder:2005vt}%
  \BibitemOpen
  \bibfield  {author} {\bibinfo {author} {\bibfnamefont {D.}~\bibnamefont
  {Roder}}, \bibinfo {author} {\bibfnamefont {J.}~\bibnamefont {Ruppert}}, \
  and\ \bibinfo {author} {\bibfnamefont {D.~H.}\ \bibnamefont {Rischke}},\
  }\href {\doibase 10.1016/j.nuclphysa.2006.05.007} {\bibfield  {journal}
  {\bibinfo  {journal} {Nucl. Phys.}\ }\textbf {\bibinfo {volume} {A775}},\
  \bibinfo {pages} {127} (\bibinfo {year} {2006})},\ \Eprint
  {http://arxiv.org/abs/hep-ph/0503042} {arXiv:hep-ph/0503042 [hep-ph]}
  \BibitemShut {NoStop}%
\bibitem [{\citenamefont {Ivanov}\ \emph {et~al.}(2005)\citenamefont {Ivanov},
  \citenamefont {Riek}, \citenamefont {van Hees},\ and\ \citenamefont
  {Knoll}}]{Ivanov:2005bv}%
  \BibitemOpen
  \bibfield  {author} {\bibinfo {author} {\bibfnamefont {{\relax Yu}.~B.}\
  \bibnamefont {Ivanov}}, \bibinfo {author} {\bibfnamefont {F.}~\bibnamefont
  {Riek}}, \bibinfo {author} {\bibfnamefont {H.}~\bibnamefont {van Hees}}, \
  and\ \bibinfo {author} {\bibfnamefont {J.}~\bibnamefont {Knoll}},\ }\href
  {\doibase 10.1103/PhysRevD.72.036008} {\bibfield  {journal} {\bibinfo
  {journal} {Phys. Rev.}\ }\textbf {\bibinfo {volume} {D72}},\ \bibinfo {pages}
  {036008} (\bibinfo {year} {2005})},\ \Eprint
  {http://arxiv.org/abs/hep-ph/0506157} {arXiv:hep-ph/0506157 [hep-ph]}
  \BibitemShut {NoStop}%
\bibitem [{\citenamefont {Seel}\ \emph {et~al.}(2012)\citenamefont {Seel},
  \citenamefont {Struber}, \citenamefont {Giacosa},\ and\ \citenamefont
  {Rischke}}]{Seel:2011ju}%
  \BibitemOpen
  \bibfield  {author} {\bibinfo {author} {\bibfnamefont {E.}~\bibnamefont
  {Seel}}, \bibinfo {author} {\bibfnamefont {S.}~\bibnamefont {Struber}},
  \bibinfo {author} {\bibfnamefont {F.}~\bibnamefont {Giacosa}}, \ and\
  \bibinfo {author} {\bibfnamefont {D.~H.}\ \bibnamefont {Rischke}},\ }\href
  {\doibase 10.1103/PhysRevD.86.125010} {\bibfield  {journal} {\bibinfo
  {journal} {Phys. Rev.}\ }\textbf {\bibinfo {volume} {D86}},\ \bibinfo {pages}
  {125010} (\bibinfo {year} {2012})},\ \Eprint {http://arxiv.org/abs/1108.1918}
  {arXiv:1108.1918 [hep-ph]} \BibitemShut {NoStop}%
\bibitem [{\citenamefont {Mao}(2014)}]{Mao:2013gva}%
  \BibitemOpen
  \bibfield  {author} {\bibinfo {author} {\bibfnamefont {H.}~\bibnamefont
  {Mao}},\ }\href {\doibase 10.1016/j.nuclphysa.2014.02.011} {\bibfield
  {journal} {\bibinfo  {journal} {Nucl. Phys.}\ }\textbf {\bibinfo {volume}
  {A925}},\ \bibinfo {pages} {185} (\bibinfo {year} {2014})},\ \Eprint
  {http://arxiv.org/abs/1305.4329} {arXiv:1305.4329 [hep-ph]} \BibitemShut
  {NoStop}%
\bibitem [{\citenamefont {Mark\'{o}}\ \emph {et~al.}(2013)\citenamefont
  {Mark\'{o}}, \citenamefont {Reinosa},\ and\ \citenamefont
  {Sz\'{e}p}}]{Marko:2013lxa}%
  \BibitemOpen
  \bibfield  {author} {\bibinfo {author} {\bibfnamefont {G.}~\bibnamefont
  {Mark\'{o}}}, \bibinfo {author} {\bibfnamefont {U.}~\bibnamefont {Reinosa}},
  \ and\ \bibinfo {author} {\bibfnamefont {Z.}~\bibnamefont {Sz\'{e}p}},\
  }\href {\doibase 10.1103/PhysRevD.87.105001} {\bibfield  {journal} {\bibinfo
  {journal} {Phys.Rev.}\ }\textbf {\bibinfo {volume} {D87}},\ \bibinfo {pages}
  {105001} (\bibinfo {year} {2013})},\ \Eprint {http://arxiv.org/abs/1303.0230}
  {arXiv:1303.0230 [hep-ph]} \BibitemShut {NoStop}%
\bibitem [{\citenamefont {Mark{\'o}}\ \emph {et~al.}(2015)\citenamefont
  {Mark{\'o}}, \citenamefont {Reinosa},\ and\ \citenamefont
  {Sz{\'e}p}}]{Marko:2015gpa}%
  \BibitemOpen
  \bibfield  {author} {\bibinfo {author} {\bibfnamefont {G.}~\bibnamefont
  {Mark{\'o}}}, \bibinfo {author} {\bibfnamefont {U.}~\bibnamefont {Reinosa}},
  \ and\ \bibinfo {author} {\bibfnamefont {Z.}~\bibnamefont {Sz{\'e}p}},\
  }\href {\doibase 10.1103/PhysRevD.92.125035} {\bibfield  {journal} {\bibinfo
  {journal} {Phys. Rev.}\ }\textbf {\bibinfo {volume} {D92}},\ \bibinfo {pages}
  {125035} (\bibinfo {year} {2015})},\ \Eprint
  {http://arxiv.org/abs/1510.04932} {arXiv:1510.04932 [hep-ph]} \BibitemShut
  {NoStop}%
\bibitem [{\citenamefont {Engels}\ and\ \citenamefont
  {Vogt}(2010)}]{Engels:2009tv}%
  \BibitemOpen
  \bibfield  {author} {\bibinfo {author} {\bibfnamefont {J.}~\bibnamefont
  {Engels}}\ and\ \bibinfo {author} {\bibfnamefont {O.}~\bibnamefont {Vogt}},\
  }\href {\doibase 10.1016/j.nuclphysb.2010.02.006} {\bibfield  {journal}
  {\bibinfo  {journal} {Nucl. Phys.}\ }\textbf {\bibinfo {volume} {B832}},\
  \bibinfo {pages} {538} (\bibinfo {year} {2010})},\ \Eprint
  {http://arxiv.org/abs/0911.1939} {arXiv:0911.1939 [hep-lat]} \BibitemShut
  {NoStop}%
\bibitem [{\citenamefont {Berges}\ \emph {et~al.}(2002)\citenamefont {Berges},
  \citenamefont {Tetradis},\ and\ \citenamefont {Wetterich}}]{Berges:2000ew}%
  \BibitemOpen
  \bibfield  {author} {\bibinfo {author} {\bibfnamefont {J.}~\bibnamefont
  {Berges}}, \bibinfo {author} {\bibfnamefont {N.}~\bibnamefont {Tetradis}}, \
  and\ \bibinfo {author} {\bibfnamefont {C.}~\bibnamefont {Wetterich}},\ }\href
  {\doibase 10.1016/S0370-1573(01)00098-9} {\bibfield  {journal} {\bibinfo
  {journal} {Phys. Rept.}\ }\textbf {\bibinfo {volume} {363}},\ \bibinfo
  {pages} {223} (\bibinfo {year} {2002})},\ \Eprint
  {http://arxiv.org/abs/hep-ph/0005122} {arXiv:hep-ph/0005122} \BibitemShut
  {NoStop}%
\bibitem [{\citenamefont {Pawlowski}(2007)}]{Pawlowski:2005xe}%
  \BibitemOpen
  \bibfield  {author} {\bibinfo {author} {\bibfnamefont {J.~M.}\ \bibnamefont
  {Pawlowski}},\ }\href {\doibase 10.1016/j.aop.2007.01.007} {\bibfield
  {journal} {\bibinfo  {journal} {Annals Phys.}\ }\textbf {\bibinfo {volume}
  {322}},\ \bibinfo {pages} {2831} (\bibinfo {year} {2007})},\ \Eprint
  {http://arxiv.org/abs/hep-th/0512261} {arXiv:hep-th/0512261 [hep-th]}
  \BibitemShut {NoStop}%
\bibitem [{\citenamefont {Gies}(2012)}]{Gies:2006wv}%
  \BibitemOpen
  \bibfield  {author} {\bibinfo {author} {\bibfnamefont {H.}~\bibnamefont
  {Gies}},\ }\href {\doibase 10.1007/978-3-642-27320-9_6} {\bibfield  {journal}
  {\bibinfo  {journal} {Lect.Notes Phys.}\ }\textbf {\bibinfo {volume} {852}},\
  \bibinfo {pages} {287} (\bibinfo {year} {2012})},\ \Eprint
  {http://arxiv.org/abs/hep-ph/0611146} {arXiv:hep-ph/0611146 [hep-ph]}
  \BibitemShut {NoStop}%
\bibitem [{\citenamefont {Schaefer}\ and\ \citenamefont
  {Wambach}(2008)}]{Schaefer:2006sr}%
  \BibitemOpen
  \bibfield  {author} {\bibinfo {author} {\bibfnamefont {B.-J.}\ \bibnamefont
  {Schaefer}}\ and\ \bibinfo {author} {\bibfnamefont {J.}~\bibnamefont
  {Wambach}},\ }\href {\doibase 10.1134/S1063779608070083} {\bibfield
  {journal} {\bibinfo  {journal} {Phys.Part.Nucl.}\ }\textbf {\bibinfo {volume}
  {39}},\ \bibinfo {pages} {1025} (\bibinfo {year} {2008})},\ \Eprint
  {http://arxiv.org/abs/hep-ph/0611191} {arXiv:hep-ph/0611191 [hep-ph]}
  \BibitemShut {NoStop}%
\bibitem [{\citenamefont {Braun}(2012)}]{Braun:2011pp}%
  \BibitemOpen
  \bibfield  {author} {\bibinfo {author} {\bibfnamefont {J.}~\bibnamefont
  {Braun}},\ }\href {\doibase 10.1088/0954-3899/39/3/033001} {\bibfield
  {journal} {\bibinfo  {journal} {J.Phys.}\ }\textbf {\bibinfo {volume}
  {G39}},\ \bibinfo {pages} {033001} (\bibinfo {year} {2012})},\ \Eprint
  {http://arxiv.org/abs/1108.4449} {arXiv:1108.4449 [hep-ph]} \BibitemShut
  {NoStop}%
\bibitem [{\citenamefont {Canet}\ \emph
  {et~al.}(2003{\natexlab{a}})\citenamefont {Canet}, \citenamefont {Delamotte},
  \citenamefont {Mouhanna},\ and\ \citenamefont {Vidal}}]{Canet:2002gs}%
  \BibitemOpen
  \bibfield  {author} {\bibinfo {author} {\bibfnamefont {L.}~\bibnamefont
  {Canet}}, \bibinfo {author} {\bibfnamefont {B.}~\bibnamefont {Delamotte}},
  \bibinfo {author} {\bibfnamefont {D.}~\bibnamefont {Mouhanna}}, \ and\
  \bibinfo {author} {\bibfnamefont {J.}~\bibnamefont {Vidal}},\ }\href
  {\doibase 10.1103/PhysRevD.67.065004} {\bibfield  {journal} {\bibinfo
  {journal} {Phys. Rev.}\ }\textbf {\bibinfo {volume} {D67}},\ \bibinfo {pages}
  {065004} (\bibinfo {year} {2003}{\natexlab{a}})},\ \Eprint
  {http://arxiv.org/abs/hep-th/0211055} {arXiv:hep-th/0211055 [hep-th]}
  \BibitemShut {NoStop}%
\bibitem [{\citenamefont {Canet}\ \emph
  {et~al.}(2003{\natexlab{b}})\citenamefont {Canet}, \citenamefont {Delamotte},
  \citenamefont {Mouhanna},\ and\ \citenamefont {Vidal}}]{Canet:2003qd}%
  \BibitemOpen
  \bibfield  {author} {\bibinfo {author} {\bibfnamefont {L.}~\bibnamefont
  {Canet}}, \bibinfo {author} {\bibfnamefont {B.}~\bibnamefont {Delamotte}},
  \bibinfo {author} {\bibfnamefont {D.}~\bibnamefont {Mouhanna}}, \ and\
  \bibinfo {author} {\bibfnamefont {J.}~\bibnamefont {Vidal}},\ }\href
  {\doibase 10.1103/PhysRevB.68.064421} {\bibfield  {journal} {\bibinfo
  {journal} {Phys.Rev.}\ }\textbf {\bibinfo {volume} {B68}},\ \bibinfo {pages}
  {064421} (\bibinfo {year} {2003}{\natexlab{b}})},\ \Eprint
  {http://arxiv.org/abs/hep-th/0302227} {arXiv:hep-th/0302227 [hep-th]}
  \BibitemShut {NoStop}%
\bibitem [{\citenamefont {Helmboldt}\ \emph {et~al.}(2015)\citenamefont
  {Helmboldt}, \citenamefont {Pawlowski},\ and\ \citenamefont
  {Strodthoff}}]{Helmboldt:2014iya}%
  \BibitemOpen
  \bibfield  {author} {\bibinfo {author} {\bibfnamefont {A.~J.}\ \bibnamefont
  {Helmboldt}}, \bibinfo {author} {\bibfnamefont {J.~M.}\ \bibnamefont
  {Pawlowski}}, \ and\ \bibinfo {author} {\bibfnamefont {N.}~\bibnamefont
  {Strodthoff}},\ }\href {\doibase 10.1103/PhysRevD.91.054010} {\bibfield
  {journal} {\bibinfo  {journal} {Phys.Rev.}\ }\textbf {\bibinfo {volume}
  {D91}},\ \bibinfo {pages} {054010} (\bibinfo {year} {2015})},\ \Eprint
  {http://arxiv.org/abs/1409.8414} {arXiv:1409.8414 [hep-ph]} \BibitemShut
  {NoStop}%
\bibitem [{\citenamefont {Wetterich}(1993{\natexlab{b}})}]{Wetterich:1991be}%
  \BibitemOpen
  \bibfield  {author} {\bibinfo {author} {\bibfnamefont {C.}~\bibnamefont
  {Wetterich}},\ }\href {\doibase 10.1007/BF01474340} {\bibfield  {journal}
  {\bibinfo  {journal} {Z.Phys.}\ }\textbf {\bibinfo {volume} {C57}},\ \bibinfo
  {pages} {451} (\bibinfo {year} {1993}{\natexlab{b}})}\BibitemShut {NoStop}%
\bibitem [{\citenamefont {{Pawlowski}}\ and\ \citenamefont
  {{Rennecke}}(2014)}]{Pawlowski:2014zaa}%
  \BibitemOpen
  \bibfield  {author} {\bibinfo {author} {\bibfnamefont {J.~M.}\ \bibnamefont
  {{Pawlowski}}}\ and\ \bibinfo {author} {\bibfnamefont {F.}~\bibnamefont
  {{Rennecke}}},\ }\href {\doibase 10.1103/PhysRevD.90.076002} {\bibfield
  {journal} {\bibinfo  {journal} {\prd}\ }\textbf {\bibinfo {volume} {90}},\
  \bibinfo {eid} {076002} (\bibinfo {year} {2014})},\ \Eprint
  {http://arxiv.org/abs/1403.1179} {arXiv:1403.1179 [hep-ph]} \BibitemShut
  {NoStop}%
\bibitem [{\citenamefont {Huber}\ and\ \citenamefont
  {Braun}(2012)}]{Huber:2011qr}%
  \BibitemOpen
  \bibfield  {author} {\bibinfo {author} {\bibfnamefont {M.~Q.}\ \bibnamefont
  {Huber}}\ and\ \bibinfo {author} {\bibfnamefont {J.}~\bibnamefont {Braun}},\
  }\href {\doibase 10.1016/j.cpc.2012.01.014} {\bibfield  {journal} {\bibinfo
  {journal} {Comput.Phys.Commun.}\ }\textbf {\bibinfo {volume} {183}},\
  \bibinfo {pages} {1290} (\bibinfo {year} {2012})},\ \Eprint
  {http://arxiv.org/abs/1102.5307} {arXiv:1102.5307 [hep-th]} \BibitemShut
  {NoStop}%
\bibitem [{\citenamefont {Cyrol}\ \emph
  {et~al.}(2016{\natexlab{b}})\citenamefont {Cyrol}, \citenamefont {Mitter},\
  and\ \citenamefont {Strodthoff}}]{Cyrol:2016zqb}%
  \BibitemOpen
  \bibfield  {author} {\bibinfo {author} {\bibfnamefont {A.~K.}\ \bibnamefont
  {Cyrol}}, \bibinfo {author} {\bibfnamefont {M.}~\bibnamefont {Mitter}}, \
  and\ \bibinfo {author} {\bibfnamefont {N.}~\bibnamefont {Strodthoff}},\
  }\href@noop {} {\  (\bibinfo {year} {2016}{\natexlab{b}})},\ \Eprint
  {http://arxiv.org/abs/1610.09331} {arXiv:1610.09331 [hep-ph]} \BibitemShut
  {NoStop}%
\bibitem [{\citenamefont {Kuipers}\ \emph {et~al.}(2013)\citenamefont
  {Kuipers}, \citenamefont {Ueda}, \citenamefont {Vermaseren},\ and\
  \citenamefont {Vollinga}}]{Kuipers:2012rf}%
  \BibitemOpen
  \bibfield  {author} {\bibinfo {author} {\bibfnamefont {J.}~\bibnamefont
  {Kuipers}}, \bibinfo {author} {\bibfnamefont {T.}~\bibnamefont {Ueda}},
  \bibinfo {author} {\bibfnamefont {J.~A.~M.}\ \bibnamefont {Vermaseren}}, \
  and\ \bibinfo {author} {\bibfnamefont {J.}~\bibnamefont {Vollinga}},\ }\href
  {\doibase 10.1016/j.cpc.2012.12.028} {\bibfield  {journal} {\bibinfo
  {journal} {Comput. Phys. Commun.}\ }\textbf {\bibinfo {volume} {184}},\
  \bibinfo {pages} {1453} (\bibinfo {year} {2013})},\ \Eprint
  {http://arxiv.org/abs/1203.6543} {arXiv:1203.6543 [cs.SC]} \BibitemShut
  {NoStop}%
\bibitem [{\citenamefont {Pawlowski}\ \emph {et~al.}(2016)\citenamefont
  {Pawlowski}, \citenamefont {Scherzer}, \citenamefont {Strodthoff},\ and\
  \citenamefont {Wink}}]{PSSW}%
  \BibitemOpen
  \bibfield  {author} {\bibinfo {author} {\bibfnamefont {J.~M.}\ \bibnamefont
  {Pawlowski}}, \bibinfo {author} {\bibfnamefont {M.}~\bibnamefont {Scherzer}},
  \bibinfo {author} {\bibfnamefont {N.}~\bibnamefont {Strodthoff}}, \ and\
  \bibinfo {author} {\bibfnamefont {N.}~\bibnamefont {Wink}},\ }\href@noop {}
  {\bibfield  {journal} {\bibinfo  {journal} {in preparation}\ } (\bibinfo
  {year} {2016})}\BibitemShut {NoStop}%
\bibitem [{\citenamefont {Carrington}\ \emph {et~al.}(2014)\citenamefont
  {Carrington}, \citenamefont {Fu}, \citenamefont {Mikula},\ and\ \citenamefont
  {Pickering}}]{Carrington:2013jta}%
  \BibitemOpen
  \bibfield  {author} {\bibinfo {author} {\bibfnamefont {M.}~\bibnamefont
  {Carrington}}, \bibinfo {author} {\bibfnamefont {W.-J.}\ \bibnamefont {Fu}},
  \bibinfo {author} {\bibfnamefont {P.}~\bibnamefont {Mikula}}, \ and\ \bibinfo
  {author} {\bibfnamefont {D.}~\bibnamefont {Pickering}},\ }\href {\doibase
  10.1103/PhysRevD.89.025013} {\bibfield  {journal} {\bibinfo  {journal}
  {Phys.Rev.}\ }\textbf {\bibinfo {volume} {D89}},\ \bibinfo {pages} {025013}
  (\bibinfo {year} {2014})},\ \Eprint {http://arxiv.org/abs/1310.4352}
  {arXiv:1310.4352 [hep-ph]} \BibitemShut {NoStop}%
\bibitem [{\citenamefont {Rose}\ \emph {et~al.}(2016)\citenamefont {Rose},
  \citenamefont {Benitez}, \citenamefont {L{\'e}onard},\ and\ \citenamefont
  {Delamotte}}]{Rose:2016wqz}%
  \BibitemOpen
  \bibfield  {author} {\bibinfo {author} {\bibfnamefont {F.}~\bibnamefont
  {Rose}}, \bibinfo {author} {\bibfnamefont {F.}~\bibnamefont {Benitez}},
  \bibinfo {author} {\bibfnamefont {F.}~\bibnamefont {L{\'e}onard}}, \ and\
  \bibinfo {author} {\bibfnamefont {B.}~\bibnamefont {Delamotte}},\ }\href
  {\doibase 10.1103/PhysRevD.93.125018} {\bibfield  {journal} {\bibinfo
  {journal} {Phys. Rev.}\ }\textbf {\bibinfo {volume} {D93}},\ \bibinfo {pages}
  {125018} (\bibinfo {year} {2016})},\ \Eprint
  {http://arxiv.org/abs/1604.05285} {arXiv:1604.05285 [cond-mat.stat-mech]}
  \BibitemShut {NoStop}%
\bibitem [{\citenamefont {Schmidt}\ and\ \citenamefont
  {Enss}(2011)}]{Schmidt:2011zu}%
  \BibitemOpen
  \bibfield  {author} {\bibinfo {author} {\bibfnamefont {R.}~\bibnamefont
  {Schmidt}}\ and\ \bibinfo {author} {\bibfnamefont {T.}~\bibnamefont {Enss}},\
  }\href {\doibase 10.1103/PhysRevA.83.063620} {\bibfield  {journal} {\bibinfo
  {journal} {Phys. Rev.}\ }\textbf {\bibinfo {volume} {A83}},\ \bibinfo {pages}
  {063620} (\bibinfo {year} {2011})},\ \Eprint {http://arxiv.org/abs/1104.1379}
  {arXiv:1104.1379 [cond-mat.quant-gas]} \BibitemShut {NoStop}%
\bibitem [{ode()}]{odeboost}%
  \BibitemOpen
  \href@noop {} {}\bibinfo {note} {K. Ahnert and M. Mulansky,
  \url{http://www.boost.org/doc/libs/1_60_0/libs/numeric/odeint/doc/html/index.html}}\BibitemShut
  {NoStop}%
\end{thebibliography}%
\end{document}